\documentclass[onecolumn,showpacs,preprintnumbers,nofootinbib,prl,superscriptaddress,groupedaddress,10pt,aps]{article}
\usepackage{graphicx,amssymb,amsmath,amsthm,amsfonts,epsfig,epsf}
\usepackage[usenames]{color}
\usepackage[linktocpage]{hyperref}
\usepackage{epstopdf,pifont}
\usepackage{cite}

\usepackage{bm}
\usepackage{dcolumn}
\usepackage[latin1]{inputenc}
\usepackage{latexsym}
\usepackage{rotating}
\usepackage{hyperref}
\usepackage{color}
\usepackage{longtable}
\usepackage{enumerate}
\usepackage{tensor}
\usepackage{multirow}
\usepackage{url}
\usepackage{authblk}
\usepackage{gensymb}
\newcommand{\yes}{\checkmark}
\newcommand{\no}{\ding{55}}

\newcommand{\ben}{\begin{enumerate}}
\newcommand{\een}{\end{enumerate}}

\def\be{\begin{equation}}
\def\ee{\end{equation}}
\def\bea{\begin{eqnarray}}
\def\eea{\end{eqnarray}}

\newcommand{\beq}{\begin{eqnarray}}
\newcommand{\eeq}{\end{eqnarray}} 
\newcommand{\ba}{\begin{align}}
\newcommand{\ea}{\end{align}}

\begin{document}

\title{The observational evidence for horizons:\\
from echoes to precision gravitational-wave physics}

\author[1,2]{Vitor Cardoso}
\author[3,1]{Paolo Pani}

\affil[1]{CENTRA, Departamento de F\'{\i}sica, Instituto Superior T\'ecnico, Universidade de Lisboa, Avenida~Rovisco Pais 1, 1049 Lisboa, Portugal}
\affil[2]{Perimeter Institute for Theoretical Physics, 31 Caroline Street North Waterloo, Ontario N2L 2Y5, Canada}
\affil[3]{Dipartimento di Fisica, ``Sapienza'' Universit\`a di Roma \& Sezione INFN Roma1, Piazzale Aldo Moro 5, 00185, Roma, Italy}

\maketitle

\begin{abstract}
The existence of black holes and of spacetime singularities is a fundamental issue in science. Despite this, observations supporting their existence are scarce,
and their interpretation unclear. We overview how strong a case for black holes has been made in the last few decades, and how well observations adjust to this paradigm.
Unsurprisingly, we conclude that observational proof for black holes is impossible to come by. However, just like Popper's black swan, alternatives can be ruled out or confirmed to exist with a single observation. These observations are within reach. In the next few years and decades, we will enter the era of precision gravitational-wave physics
with more sensitive detectors. Just as accelerators require larger and larger energies to probe smaller and smaller scales, more 
sensitive gravitational-wave detectors will be probing regions closer and closer to the horizon, potentially reaching Planck scales and beyond. What may be there, lurking?\end{abstract}

\newpage

\tableofcontents

\newpage
\hspace{1.8cm}
\parbox{0.8\textwidth}{{\small 
\noindent {\it ``The crushing of matter to infinite density by infinite tidal gravitation forces is a phenomenon with which one cannot live comfortably.
From a purely philosophical standpoint it is difficult to believe that physical singularities are a fundamental and unavoidable feature of our universe
[...] one is inclined to discard or modify that theory rather than accept the suggestion that the singularity actually occurs in nature.''}
% \vskip 2mm
\begin{flushright}
Kip Thorne, Relativistic Stellar Structure and Dynamics (1966) 
\end{flushright}
}
}
\vskip 1cm
\hspace{1.8cm}
\parbox{0.8\textwidth}{{\small 
\noindent {\it ``No testimony is sufficient to establish a miracle, unless the testimony be of such a kind, that its falsehood would be more miraculous than the fact which it endeavors to establish.''}
% \vskip 2mm
\begin{flushright}
David Hume, An Enquiry concerning Human Understanding (1748) 
\end{flushright}
}
}

\vskip 1cm

%%%%%%%%%%%%%%%%%%%%%%%%%%%%%%%%%%%%%%%%%%%%%%%%%%%%%%%%%%%%%%%%%%%%%%%%%%%%%%
\section{Introduction}
%%%%%%%%%%%%%%%%%%%%%%%%%%%%%%%%%%%%%%%%%%%%%%%%%%%%%%%%%%%%%%%%%%%%%%%%%%%%%%

The discovery of the electron and the known neutrality of matter led in 1904 to J. J. Thomson's ``plum-pudding'' atomic model. Data
from new scattering experiments was soon found to be in tension with this model, which was eventually superseeded by Rutherford's, featuring an atomic nucleus.
The point-like character of elementary particles opened up new questions. How to explain the apparent stability of the atom? How to handle the singular behavior of
the electric field close to the source? What is the structure of elementary particles? Some of these questions were elucidated with quantum mechanics and quantum field theory. Invariably, the path to the answer led to the understanding of hitherto unknown phenomena.

The history of elementary particles is a timeline of the understanding of the electromagnetic (EM) interaction, and is pegged to its characteristic $1/r^2$ behavior (which necessarily implies that other structure {\it has} to exist on small scales within any sound theory).

Arguably, the elementary particle of the gravitational interaction are black holes (BHs). Within General Relativity (GR), BHs are indivisible and are in fact the simplest macroscopic objects that one can conceive. The uniqueness theorems --~establishing that the two-parameter Kerr family of BHs describes any vacuum, stationary and asymptotically flat, regular BH solution~--
have turned BHs into somewhat of a miracle elementary particle~\cite{Chandra}.

Can BHs, as envisioned in vacuum GR, hold the same surprises that the electron and the hydrogen atom did when they started to be experimentally probed?
Are there any parallels that can be useful guides? 
The BH interior is causally disconnected from the exterior by an event horizon. Unlike the classical description of atoms, the GR description of the BH exterior is self-consistent and free of pathologies. 
The ``inverse-square law problem'' --~the GR counterpart of which is the appearance of pathological curvature singularities~-- is swept to inside the horizon and therefore harmless for the external world. 
There are good indications that classical BHs are perturbatively stable against small fluctuations, and attempts to produce naked singularities, starting from BH spacetimes, have failed. BHs are not only curious mathematical solutions to Einstein's equations, but also their {\it formation} process is well understood.
In fact, there is nothing spectacular with the presence or formation of an event horizon. The equivalence principle dictates that an infalling observer crossing this region (which, by definition, is a \emph{global} concept)
feels nothing extraordinary: in the case of macroscopic BHs 
all of the local physics at the horizon is rather unremarkable.

Why, then, should one question the existence of BHs?~\footnote{%Given the history of these objects, 
It is ironic that one asks this question more than a century after the first BH solution was derived. 
Even though Schwarzschild and Droste wrote down the first nontrivial regular, asymptotic flat, vacuum solution to the field equations already in 1916~\cite{Schwarzschild:1916uq,Droste:1916uq}, several decades would elapse until such solutions became accepted. 
The dissension between Eddington and Chandrasekhar over gravitational collapse to BHs is famous --~Eddington firmly believed that nature would find its way to prevent full collapse~-- and it took decades for the community to overcome individual prejudices.
Progress in our understanding of gravitational collapse, along with breakthroughs in the understanding of singularities and horizons, contributed to change the status of BHs. 
Together with observations of phenomena so powerful that could only be explained via massive compact objects, the theoretical understanding of BHs turned them into undisputed kings of the cosmos.
The irony lies in the fact that they quickly became the {\it only} acceptable solution. So much so, that currently an informal definition of a BH might well be ``any dark, compact object with mass above three solar masses.''} 
There are a number of important reasons to do so.

\begin{itemize}

\item Even if the BH exterior is pathology-free, the interior is not. The Kerr family of BHs harbors, singularities and closed timelike curves in its interior, and more generically it features a Cauchy horizon signaling the breakdown of predictability of the theory. In fact, the geometry describing the interior of an astrophysical spinning BH is currently unknown. A possible resolution of this problem may require accounting for quantum effects. It is conceivable that these quantum effects are of no consequence whatsoever on physics outside the horizon. Nevertheless, it is conceivable as well that the resolution of such inconsistency leads to new physics that resolves singularities and does away with horizons, at least in the way we understand them currently. Such possibility is not too dissimilar from what happened with the atomic model after the advent of quantum electrodynamics.

\item In a related vein, quantum effects around BHs are far from being under control, with the evaporation process itself leading to information loss. The resolution of such problems could include changing the endstate of collapse~\cite{Mazur:2004fk,Mathur:2005zp,Mathur:2008nj,Barcelo:2015noa,Kawai:2017txu} (perhaps as-yet-unknown physics can prevent the formation of horizons), 
or altering drastically the near-horizon region~\cite{Unruh:2017uaw}. 
The fact of the matter is that there is no tested nor fully satisfactory theory of quantum gravity, in much the same way that one did not have a quantum theory of point particles at the beginning of the 20th century.

\item It is tacitly assumed that quantum gravity effects become important only at the Planck scale. At such lengthscales $\sqrt{c \hbar/G}$, the Schwarzschild radius is of the order of the 
Compton wavelength of the BH. In the orders of magnitude standing between the Planck scale and those accessible by current experiments many surprises can hide
(to give but one example, extra dimensions would change the numerical value of the Planck scale).

\item Horizons are not only a rather generic prediction of GR, but their existence is in fact \emph{necessary} for the consistency of the theory at the classical level. This is the root of Penrose's (weak) Cosmic Censorship Conjecture, which remains one of the most urgent open problems in fundamental physics. In this sense, the statement that there is a horizon in any spacetime harboring a singularity in its interior is such a remarkable claim, that (in an informal description of Hume's statement above) it requires similar remarkable evidence.
This is the question we will entertain here, {\it is there any observational evidence for the existence of BHs?}

\item It is in the nature of science that paradigms have to be constantly questioned and subjected to experimental and observational scrutiny. Most specially because if the answer turns out to be that BHs do not exist, the consequences are so extreme and profound, that it is worth all the possible burden of actually testing it. Within the coming years we will finally be in the position of performing unprecedented tests on the nature of compact dark objects, the potential pay-off of which is enormous. 
As we will argue, the question is not just whether the strong-field gravity region near compact objects is consistent with the Kerr geometry, but rather to \emph{quantify} the limits of observations in testing the event horizon.

\item If we don't entertain alternatives, we won't find them. We need to know how to model alternatives to understand how consistent
the data is with accepted paradigms.

\item Finally, there are sociological issues, sometimes invoked to claim that one should invest taxpayers money in sure-fire science.
The reality is that taxpayers, all of us, {\it want to know}. More frequently than not, the thrill lies in the intellectual possibilities rather than on merely checking known facts. This is, after all, why pure science is relevant. Thus, it is a waste of resources {\it not to use} (built and already paid for) detectors to investigate issues that lie at the heart of fundamental questions.
\end{itemize}

Known physics all but exclude BH alternatives. Nonetheless, the Standard Model of fundamental interactions
is not sufficient to describe the cosmos --~at least on the largest scales~-- and also leaves all of the fundamental questions regarding BHs open. It may be wise to keep an open mind. As we will argue, from the point of view of observers merely {\it testing} the spacetime structure without {\it a priori} theoretical bias, the evidence for horizons is nonexistent. 
In fact, the question to be asked is not if there is an event horizon in the spacetime, but how close to it do experiments or observations go.

\clearpage
\newpage
%%%%%%%%%%%%%%%%%%%%%%%%%%%%%%%%%%%%%%%%%%%%%%%%%%%%%%%%%%%%%%%%%%%%%%%%%%%%%%%%%%%%%%%%%%%%%%%%%%%%%%%%%%%%%%
\section{Setting the stage: escape cone, photospheres, quasinormal modes, and tidal effects\label{sec:stage}}
%%%%%%%%%%%%%%%%%%%%%%%%%%%%%%%%%%%%%%%%%%%%%%%%%%%%%%%%%%%%%%%%%%%%%%%%%%%%%%%%%%%%%%%%%%%%%%%%%%%%%%%%%%%%%%

\begin{flushright}
{\small 
\noindent {\it ``Alas, I abhor informality.''}\\
% \vskip 2mm
That Mitchell and Webb Look, Episode 2
}
\end{flushright}

\vskip 1cm

Our framework will be that of GR. Let us start by focusing on spherical symmetry for simplicity. Birkhoff's theorem guarantees that 
any vacuum spacetime is described by the Schwarzschild geometry, which in standard coordinates reads
\begin{equation}
ds^2=-f dt^2+f^{-1}dr^2+r^2d\Omega^2\,, \label{metric}
\end{equation}
where $M$ is the total mass of the spacetime (we use geometrical $G=c=1$ units, except if otherwise stated) and
\be
f=1-\frac{2M}{r}\,.
\ee
The existence of a null hypersurface at $r=2M$ is the defining feature of a BH. This hypersurface is called the event horizon of a Schwarzschild BH.
In the following, we will compare the properties of the latter with those of an ultracompact object with a effective surface at 
\begin{equation}
r_0=2M(1+\epsilon)\,,
\end{equation}
having often in mind the case where $\epsilon \ll 1$ (for instance, for Planckian corrections of an astrophysical BH, $\epsilon \sim 10^{-40}$). Although the above definition is coordinate-dependent, the proper distance between the surface and $r_g$ scales like $\int_{2M}^{r_0}dr f^{1/2}\sim\epsilon^{1/2}$.
Most of the results discussed below show a dependence on $\log \epsilon$, making the distinction irrelevant.
There are objects for which the effective surface is ill-defined, because the matter fields are smooth everywhere. In such cases, 
$r_0$ can conservatively be taken to be the geometric center of the object.
%%%%%%%%%%%%%%%%%%%%%%%%%%%%%%%%%%%%%%%%%%%%%%%%%%%%%%%%%%%%%%%%%%%%%%%%%%%%%%
\subsection{Geodesics}
%%%%%%%%%%%%%%%%%%%%%%%%%%%%%%%%%%%%%%%%%%%%%%%%%%%%%%%%%%%%%%%%%%%%%%%%%%%%%%

The geodesic motion of timelike or null particles in the geometry~\eqref{metric} can be described with the help of two conserved quantities,
the specific energy $E=f\,\dot{t}$ and angular momentum $L=r^2\dot{\varphi}$, where a dot stands for a derivative with respect to proper time~\cite{MTB,Cardoso:2008bp}. The radial motion can be computed via a normalization condition,
\be
\dot{r}^2=E^2-f\left(\frac{L^2}{r^2}+\delta_1\right)\equiv E^2-V_{\rm geo}\,, \label{radial_motion}
\ee
where $\delta_{1}=1,0$ for timelike or null geodesics, respectively. The null limit can be approached letting $E,L\to \infty$ and re-scaling all quantities appropriately.
At large distances, $r\gg M$, the motion of matter resembles that in Newton's theory. For instance, circular orbits, defined by $\dot{r}=\ddot{r}=0$, exist and are stable at large distance: small fluctuations in the motion restore the body to its original position.
However, circular trajectories are stable only when $r\geq6M$, and unstable for smaller radii. The $r=6M$ surface defines the innermost stable circular orbit (ISCO), and has an important role in controlling the inner part of the accretion flow onto compact objects.

%%%%%%%%%%%%%%%%%%%%%%%%%%%%%%%%%%%%%%%%%%%%%%%%%%%%%%%%%%%%%%%%%%%%%%%%%%%%%%
\subsection{Escape trajectories\label{sec:escape}}
%%%%%%%%%%%%%%%%%%%%%%%%%%%%%%%%%%%%%%%%%%%%%%%%%%%%%%%%%%%%%%%%%%%%%%%%%%%%%%
%
\begin{figure*}[ht]
% \begin{center}
\begin{tabular}{m{0.6\linewidth}m{0.4\linewidth}}
 \hspace{-0.4cm}\includegraphics[width=0.6\textwidth]{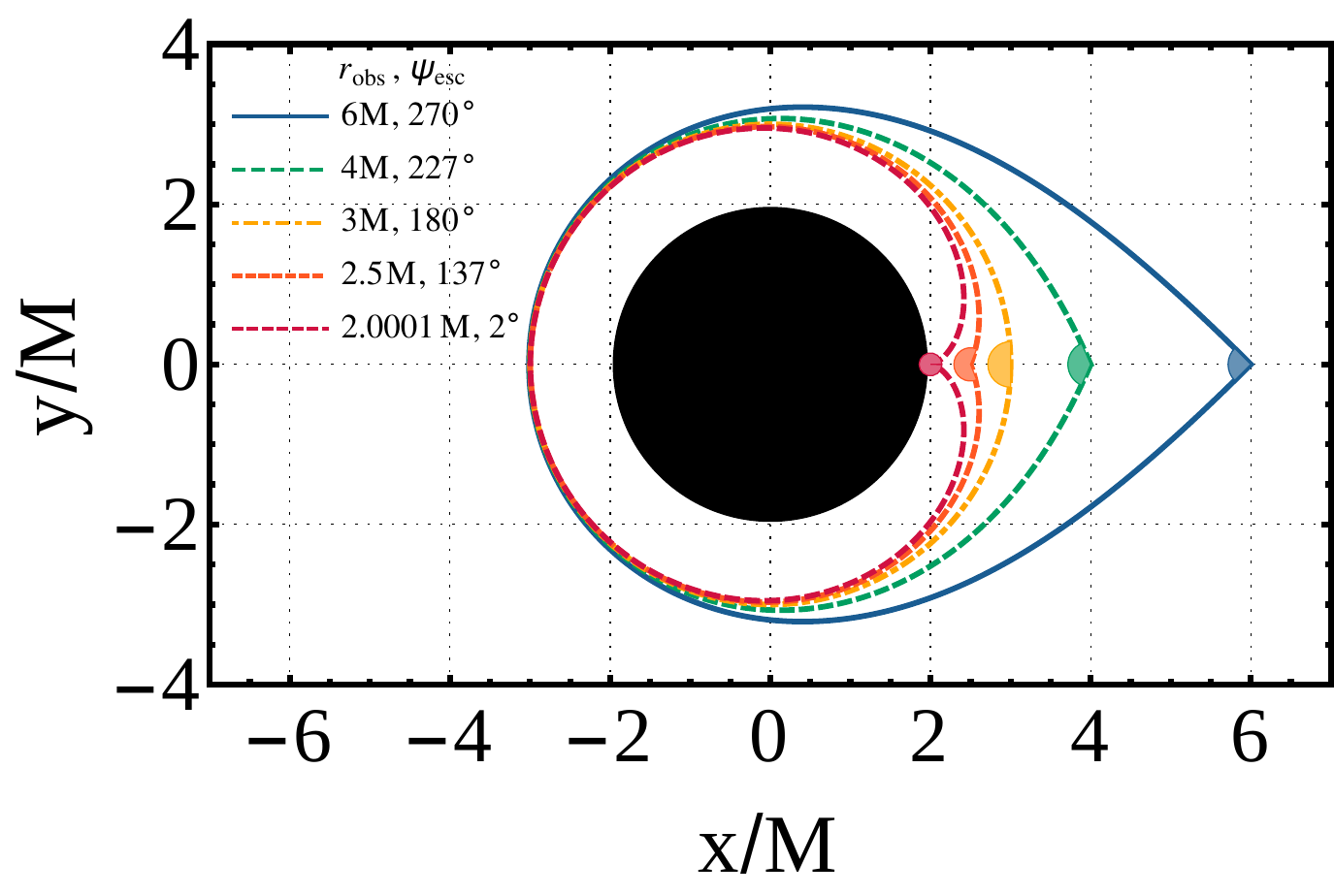} &
\hspace{-1.2cm} \includegraphics[width=0.39\textwidth]{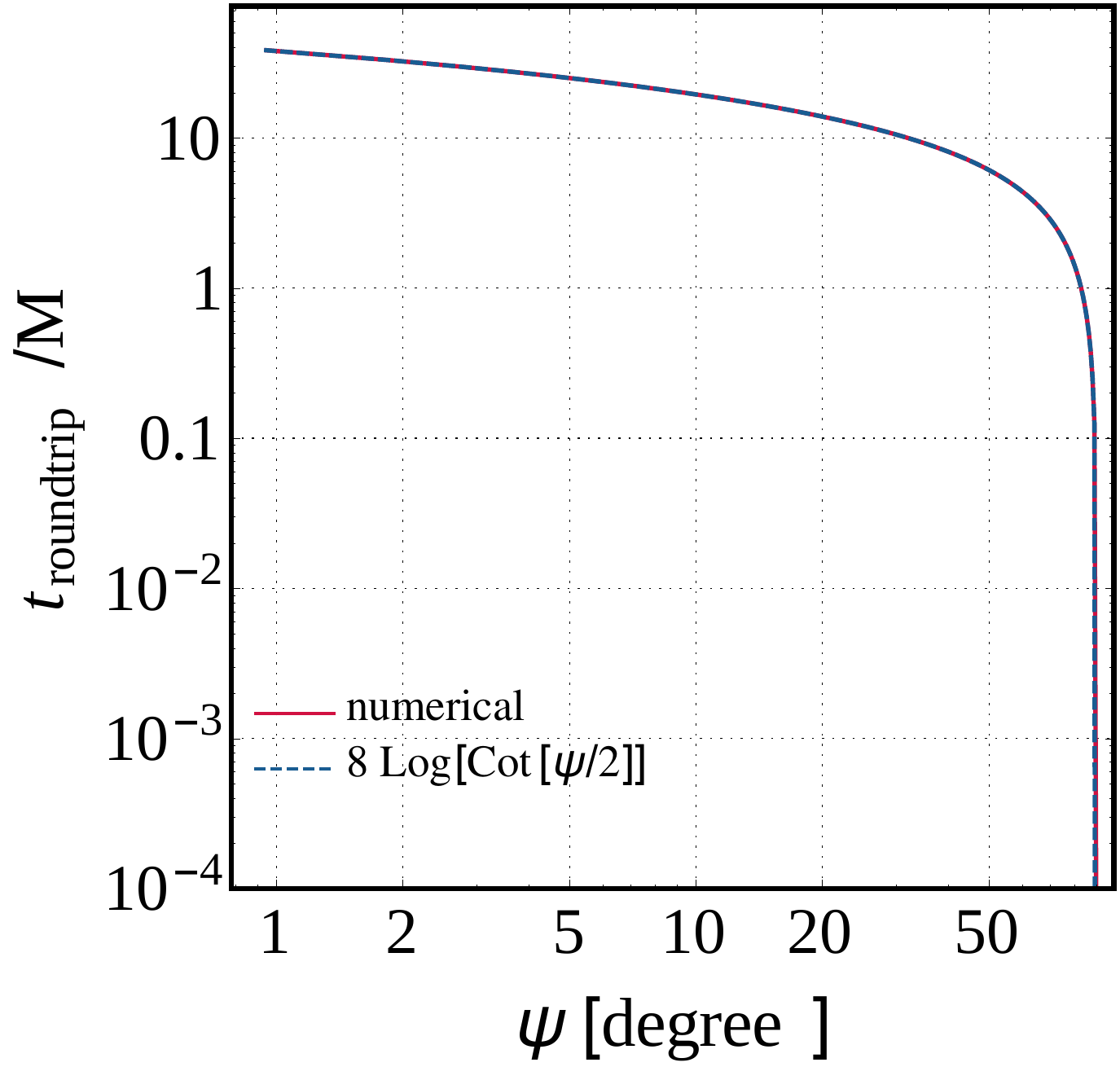}
\end{tabular}
\caption{
Left: Critical escape trajectories of radiation in the Schwarzschild geometry. A locally static observer (located at $r=r_{\rm obs}$) emits photons isotropically, but those emitted within the colored conical sectors will not reach infinity. 
\noindent Right: Coordinate roundtrip time of photons as a function of the emission angle $\psi>\psi_{\rm esc}$ and for $\epsilon \ll 1$.
\label{fig:capture}}
% \end{center}
\end{figure*}
There are other GR effects on the motion of particles around compact objects, beyond the appearance of ISCOs in spacetimes. One of them concerns the behavior of light rays and how they are bent by the spacetime curvature. There exists a critical value of the angular momentum $L\equiv KME$ for a light ray to be able to escape to infinity. By requiring that a light ray emitted at a given point 
will not find turning points in its motion, Eq.~\eqref{radial_motion} yields $K_{\rm esc}=3\sqrt{3}$~\cite{MTB}.
Suppose that the light ray is emitted by a locally static observer at $r=r_0$.
In the local rest frame, the velocity components of the photon are~\cite{Shapiro:1983du}
\beq
v_{\varphi}^{\rm local}&=&\frac{MK}{r_0}\sqrt{f_0}\,,\qquad v_{r}^{\rm local}=\sqrt{1-K^2M^2\frac{f_0}{r_0^2}}\,,
\eeq
where $f_0\equiv f(r_0)=1-2M/r_0$. 
With this, one can easily compute the escape angle, $\sin\psi_{\rm esc}=3M \sqrt{3f_0}/r_0$.
In other words, the solid angle for escape is
\begin{equation}
 \Delta \Omega_{\rm esc}= 2\pi\left(1-\sqrt{1-\frac{27 M^2(r_0-2M)}{r_0^3}}\right)\sim 27\pi\left(\frac{r_0-2M}{8M}\right)\,, \label{solidangle}
\end{equation}
where the last step is valid for $\epsilon\ll 1$.
For angles larger than these, the light ray falls back and either hits the surface of the object, if there is one, or will be absorbed by the horizon.
The escape angle is depicted in Fig.~\ref{fig:capture} for different emission points $r_0$. The rays that are not able to escape reach a maximum coordinate distance,
\be
r_{\rm max} \sim 
2M \left(1+\frac{4f_0M^2}{r_0^2\sin^2\psi}\right)\,.
\ee
This result is accurate away from $\psi_{\rm esc}$, whereas for $\psi\to\psi_{\rm esc}$ the photon approaches the photosphere ($r=3M$) discussed in the next section.
The coordinate time that it takes for photons that travel initially outward, but eventually turn back and hit the surface of the object,
is shown in Fig.~\ref{fig:capture} as a function of the locally measured angle $\psi$, and is 
of order $\sim M$ for most of the angles $\psi$, for $\epsilon\ll 1$. We find a closed form expression away from $\psi_{\rm crit}$, which describes well the full range (see Fig.~\ref{fig:capture}),
\begin{equation}
t_{\rm roundtrip}\sim 8M \log(\cot{\left(\psi/2\right)})\,. \label{troundtrip}
\end{equation}
% 
%This is an intriguing result. 
When averaging over $\psi$, the coordinate roundtrip time is then $32 M\,{\rm Cat}/\pi \approx 9.33 M$, for any $\epsilon\ll 1$, where ``Cat'' is Catalan's constant.

%%%%%%%%%%%%%%%%%%%%%%%%%%%%%%%%%%%%%%%%%%%%%%%%%%%%%%%%%%%%%%%%%%%%%%%%%%%%%%
\subsection{Photospheres, ECOs and ClePhOs}
%%%%%%%%%%%%%%%%%%%%%%%%%%%%%%%%%%%%%%%%%%%%%%%%%%%%%%%%%%%%%%%%%%%%%%%%%%%%%%
%
\begin{figure*}[ht]
\begin{center}
\includegraphics[width=1.1\textwidth]{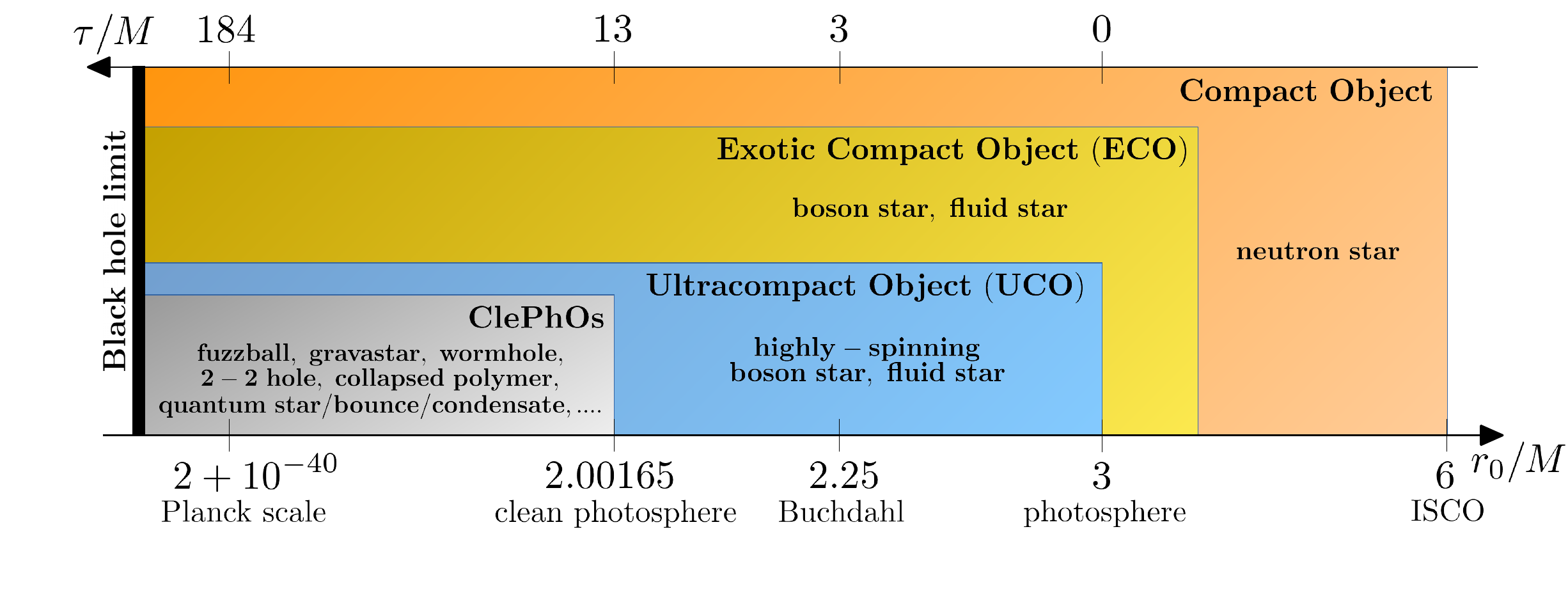}
\caption{Schematic classification of compact objects according to their compactness (in a logarithmic scale). Objects in the same category have similar dynamical properties on a timescale $\tau$. 
\label{fig:diagram}}
\end{center}
\end{figure*}
Another truly relativistic feature is the existence of circular null geodesics, i.e., of circular motion for high-frequency electromagnetic or gravitational waves (GWs). 
In the Schwarzschild geometry, this can happen only at $r=3M$. This location defines a surface called
the photosphere, or, in the context of equatorial slices, a light ring. The photosphere has a number of interesting properties, and is useful to understand certain features of spacetimes. For example, it controls how BHs look like when illuminated by accretion disks or stars, thus defining their so-called ``shadow''. Imaging these shadows for the supermassive dark source SgrA$^*$ is the main goal of the Event Horizon Telescope~\cite{Loeb:2013lfa,Goddi:2016jrs}. The photosphere also has a bearing on the spacetime response to any type of high-frequency waves, and therefore describes how high-frequency gravitational waves linger close to the horizon. 

At the photosphere, $V_{\rm geo}''=-2 E^2/(3 M^2)<0$. Thus, circular null geodesics are unstable: a displacement $\delta$ of null particles grows exponentially~\cite{Ferrari:1984zz,Cardoso:2008bp}
\begin{equation}
\delta(t) \sim \delta_0 e^{\lambda t}\,,\qquad \lambda=\sqrt{\frac{-f^2 V_r''}{2E^2}}=\frac{1}{3\sqrt{3}M}\,.
\end{equation}
A geodesic description anticipates that light or GWs may persist at or close to the photosphere on timescales 
$3\sqrt{3}M \sim 5M$. Because the geodesic calculation is local, these conclusions hold irrespectively of the spacetime being vacuum all the way to the horizon or not. An ultracompact object with surface at $r_0=2M(1+\epsilon)$,
with $\epsilon \ll 1$, would feature exactly the same geodesics and properties close to its photosphere, provided that on timescales $\sim 15M$ the null geodesics did not have time to bounce off the surface.
We are requiring three $e$-fold times for the instability to dissipate more than $99.7\%$ of the energy of the initial pulse. This amounts to requiring that
\be
\epsilon\lesssim\epsilon_{\rm crit}\sim 0.0165\,. \label{eps_crit}
\ee
Thus, the horizon plays no special role in the response of high frequency waves, nor could it: it takes an infinite (coordinate) time for a light ray to reach the horizon.
The above threshold on $\epsilon$ is a natural sifter between two classes of compact, dark objects. 
%Only those objects characterized by $\epsilon \gtrsim 0.0165$ can induce noticeable dynamics in the photosphere on sufficiently short timescales.
For objects characterized by $\epsilon \gtrsim 0.0165$, light or GWs can make the roundtrip from the photosphere to the object's surface and back, before dissipation of the photosphere modes occurs. For objects satisfying \eqref{eps_crit}, the waves trapped at the photosphere relax away by the time that the waves from the surface hit it back.

We can thus use the properties of the ISCO and photosphere to distinguish between different classes of models: an object is defined as \emph{compact} if it features an ISCO, or in other words if its 
surface satisfies $r_0<6M$. Accretion disks around compact objects of the same mass should have similar characteristics.
It is customary to define \emph{Exotic Compact Objects (ECOs)} as those horizonless compact objects more massive than a neutron star.
Among the compact objects, some feature photospheres. These could be called \emph{ultracompact objects (UCOs)}.
Finally, those objects which satisfy condition \eqref{eps_crit} have a ``clean'' photosphere, and will be designated by \emph{ClePhOs}.
The early-time dynamics of ClePhOs is expected to be the same as that of BHs. At late times, ClePhOs should display unique signatures of their surface.
The zoo of compact objects is summarized in Fig.~\ref{fig:diagram}~\footnote{This diagram can be misleading in special cases. Some objects with a surface at finite $\epsilon$ can nevertheless behave as ClePhOs. For example, constant density stars near the Buchdahl limit behave as ClePhOs for axial modes: these do not couple to the fluid and travel unimpeded to the center of the star.
Their travel time can be considerably long, hence these stars act, for all purposes, as ClePhOs for these fluctuations. A specific example is discussed in footnote 3. Nevertheless, it should be clear that these are contrived examples which are not meant to describe quantum corrections.}.

%%%%%%%%%%%%%%%%%%%%%%%%%%%%%%%%%%%%%%%%%%%%%%%%%%%%%%%%%%%%%%%%%%%%%%%%%%%%%%
\subsection{Quasinormal modes\label{sec:qnms}}
%%%%%%%%%%%%%%%%%%%%%%%%%%%%%%%%%%%%%%%%%%%%%%%%%%%%%%%%%%%%%%%%%%%%%%%%%%%%%%

Low-frequency gravitational or EM waves are not well described
by null particles. In the linearized regime, massless waves are all
described by a master partial differential equation of the form~\cite{Berti:2009kk}
\be
f^2\frac{\partial^2 \Psi(t,r)}{\partial r^2}+ff'\frac{\partial \Psi(t,r)}{\partial r}-\frac{\partial^2 \Psi(t,r)}{\partial t^2}-V(r)\Psi(t,r)=S(t,r)\,.\label{pde}
\ee
The source term $S(t,r)$ contains information about the cause of the disturbance $\Psi(t,r)$.
The information about the angular dependence of the wave is encoded in the way the separation was achieved, and involves an expansion
in scalar, vector or tensor harmonics for different spins $s$ of the perturbation. These angular functions are labeled by an integer $l>|s|$, and the effective potential is
\begin{equation}
V=f\left(\frac{l(l+1)}{r^2}+(1-s^2)\frac{2M}{r^3}\right)\,, \label{potential}
\end{equation}
with $s=0,\pm1,\pm2$ for scalar, vector or axial tensor modes. The $s=\pm2$ equation does not describe completely gravitational perturbations, the potential for polar gravitational perturbations is more complicated~\cite{Chandrasekhar:1975zza,Kokkotas:1999bd,Berti:2009kk}.

It is hard to extract general information from the PDE~\eqref{pde}, in particular because its solutions depend
on the source term and initial conditions. 
We can gain some insight by studying the source-free equation in Fourier space. By defining the Fourier transform through $\Psi(t,r)=\frac{1}{\sqrt{2\pi}}\int e^{-i\omega t}\psi(\omega,r)d\omega$, one gets the following ODE
% %
\begin{equation}
 \frac{d^2\psi}{dz^2}+\left(\omega^2-V\right)\psi=0\,, \label{ode}
\end{equation}
where $r$ is implicitly written in terms of the new ``tortoise'' coordinate
\begin{equation}
z=r+2M \log\left(\frac{r}{2M}-1\right)\,,
\end{equation}
such that $z(r)$ diverges logarithmically near the horizon.
In terms of $z$, Eq.~\eqref{ode} is equivalent to the time-independent Schr\"odinger equation in one dimension and it reduces to the wave equation governing a string when $M=l=0$. To understand a string of length $L$ with fixed ends,
one imposes Dirichlet boundary conditions and gets an eigenvalue problem for $\omega$. The boundary conditions can only be satisfied for a discrete set of \emph{normal} frequencies,  $\omega= \frac{n\pi}{L}$ ($n=1,2,...$). The corresponding wavefunctions are called normal modes and form a basis onto which one can expand any configuration of the system.
The frequency is purely real because the associated problem is conservative.

\begin{figure*}[ht]
\begin{center}
\includegraphics[width=0.8\textwidth]{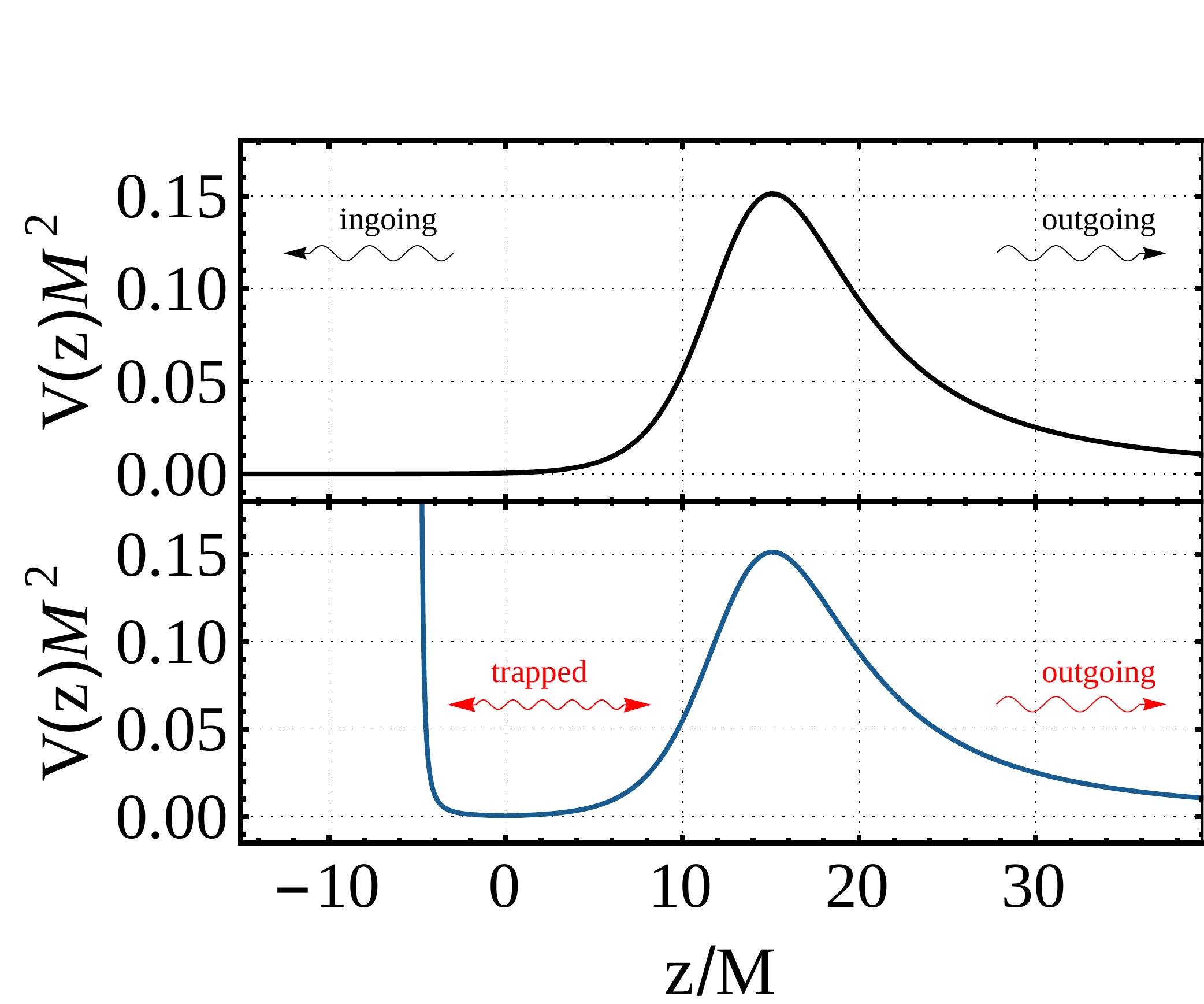}
\caption{Typical effective potential for perturbations of a Schwarzschild BH (top panel) and of an horizonless compact object (bottom panel). In the BH case, QNMs are waves which are outgoing at infinity ($z\to+\infty$) and ingoing at the horizon ($z\to-\infty$), whereas the presence of a potential well (provided either by a reflective surface, a centrifugal barrier at the center, or by the geometry) supports quasi-trapped, long-lived modes.}
\label{fig:potential}
\end{center}
\end{figure*}

If one is dealing with a BH spacetime, the appropriate conditions (required by causality) correspond to having waves traveling outward to spatial infinity ($\Psi \sim e^{i\omega z}$ as $z\to\infty$)
and inwards to the horizon ($\Psi \sim e^{-i\omega z}$ as $z\to-\infty$) [see Fig.~\ref{fig:potential}]. 
Due to backscattering off the effective potential~\eqref{potential}, the eigenvalues $\omega$ are not known in closed form,
but they can be computed numerically~\cite{Chandrasekhar:1975zza,Kokkotas:1999bd,Berti:2009kk}. The fundamental $l=2$ mode (the lowest dynamical multipole in GR) of gravitational perturbations  reads~\cite{rdweb}
\be
M\omega\equiv M(\omega_R+i\omega_I)\approx 0.373672 -i 0.0889623\,.\label{bh_qnm}
\ee
Remarkably, the entire spectrum is the same for both the axial or the polar gravitational sector~\cite{Chandrasekhar:1975zza}.
The frequencies are complex and are therefore called {\it quasinormal} frequencies. Their imaginary component describes the decay in time of fluctuations on a timescale $\tau\equiv 1/|\omega_I|$, and hints at the stability of the geometry. Unlike the case of a string with fixed end, we are now dealing with an open system:
waves can travel to infinity or down the horizon and therefore it is physically sensible that any fluctuation damps down.
The corresponding modes are the quasinormal modes (QNMs), which in general do \emph{not} form a complete set.

Boundary conditions play a crucial role in the structure of the QNM spectrum. If a reflective surface is placed at $r_0=2M(1+\epsilon)\gtrsim 2M$, where (say) Dirichlet boundary conditions have to be imposed, the spectrum changes considerably. For $\epsilon=10^{-6}$ the fundamental mode of the eigenspectrum reads
\beq
M\omega_{\rm polar}&\approx&0.13377 -i 2.8385\times 10^{-7}\\
M\omega_{\rm axial}&\approx&0.13109- i 2.3758\times 10^{-7} \,.
\eeq
Not only are the two gravitational sectors no longer isospectral but, more importantly, the perturbations have smaller frequency and are much longer lived, since a decay channel (the horizon) has disappeared. 

More generically, for Dirichlet or Neumann boundary conditions at $r=r_0$, the QNMs in the $\epsilon\to0$ limit read~\cite{Vilenkin:1978uc}
%%%%
\begin{eqnarray}
M \omega_R&\simeq&\frac{M}{2|z_0|}(p\pi-\delta)\sim |\log\epsilon|^{-1}\,, \label{omegaRecho}\\
M\omega_I &\simeq&-\beta_{ls}\frac{M}{|z_0|}(2M\omega_R)^{2l+2}\sim -|\log\epsilon|^{-(2l+3)}\,, \label{omegaIecho}
\end{eqnarray}
%%%%
where $z_0\equiv z(r_0)\sim 2M\log\epsilon$, $p$ is an odd (even) integer for Dirichlet (Neumann) boundary conditions, $\delta$ is the phase of the wave reflected at $r=r_0$, and $\beta_{ls}=\left[\frac{(l-s)!(l+s)!}{(2l)!(2l+1)!!}\right]^2$~\cite{Starobinskij2,Brito:2015oca}. The above scaling can be understood in terms of modes trapped between the peak of the potential~\eqref{potential} at $r\sim 3M$ and the ``hard surface'' at $r=r_0$~\cite{Cardoso:2016rao,Cardoso:2016oxy,Volkel:2017ofl,Mark:2017dnq} [see Fig.~\ref{fig:potential}]. 
Low-frequency waves are almost trapped by the potential, so their frequency scales as the size of the cavity (in tortoise coordinates), $\omega_R\sim1/z_0$, just like the normal modes of a string.
The (small) imaginary part is given by waves which tunnel through the potential and reach infinity. The tunneling probability can be computed analytically in the small-frequency regime and scales as $|\mathcal{A}|^2\sim (M\omega_R)^{2l+2}\ll1$~\cite{Starobinskij2}. 
After a time $t$, a wave trapped inside a box of size $z_0$ is reflected $N=t/z_0$ times, and its amplitude reduces to
$A(t)=A_0\left(1-|\mathcal{A}|^2\right)^N\sim A_0\left(1-t|\mathcal{A}|^2/z_0\right)$. 
Since, $A(t)\sim A_0 e^{-|\omega_I| t}\sim A_0(1-|\omega_I| t)$ in this limit,  we immediately obtain
%%%
\begin{equation}
 \omega_R\sim1/z_0\,,\qquad \omega_I\sim|\mathcal{A}|^2/z_0 \sim \omega_R^{2l+3}\,.
\end{equation}
%%%
This scaling agrees with exact numerical results and is valid for any $l$ and any type of perturbation.

Clearly, a perfectly reflecting surface is an idealization. In certain models, only low-frequency waves are reflected, whereas higher-frequency waves probe the internal structure of the specific object~\cite{Saravani:2012is,Mathur:2012jk}. In general, the location of the effective surface and its properties (e.g., its reflectivity) can depend on the energy scale of the process under consideration.

%%%%%%%%%%%%%%%%%%%%%%%%%%%%%%%%%%%%%%%%%%%%%%%%%%%%%%%%%%%%%%%%%%%%%%%%%%%%%%
\subsection{Quasinormal modes, photospheres, and echoes}
%%%%%%%%%%%%%%%%%%%%%%%%%%%%%%%%%%%%%%%%%%%%%%%%%%%%%%%%%%%%%%%%%%%%%%%%%%%%%%
The effective potential $V_{\rm geo}$ for geodesic motion reduces to that of wave propagation $V$
in the high-frequency, high-angular momentum (i.e., eikonal) regime. Thus, some properties of geodesic motion have a wave counterpart~\cite{Cardoso:2008bp}.
The instability of light rays along the null circular geodesic translates into some properties of waves around objects compact enough to feature a photosphere.
A wave description needs to satisfy ``quantization conditions.'' Since GWs are quadrupolar in nature, the lowest mode of vibration should satisfy 
\be
M\omega_R^{\rm geo}=2\frac{\dot{\varphi}}{\dot{t}}=\frac{2}{3\sqrt{3}}\sim 0.3849\,.
\ee
In addition the mode is damped, as we showed, on timescales $3\sqrt{3}M$. Overall then, the geodesic analysis predicts
\begin{equation}
M\omega^{\rm geo}\sim 0.3849-i0.19245\,.
\end{equation}
This crude estimate, valid in principle only for high-frequency waves, matches rather well even the fundamental mode of a Schwarzschild BH, Eq.~\eqref{bh_qnm}.

Nevertheless, QNMs frequencies can be defined for any dissipative system, not only for compact objects or BHs.
Thus, the association with photospheres has limits. Such an analogy is nonetheless enlightening
in the context of objects so compact that they have photospheres and resemble Schwarzschild deep into the geometry, in a way that condition~\eqref{eps_crit} is satisfied~\cite{Cardoso:2016rao,Cardoso:2016oxy,Price:2017cjr}.

\begin{figure*}[ht]
\begin{center}
\includegraphics[width=0.7\textwidth]{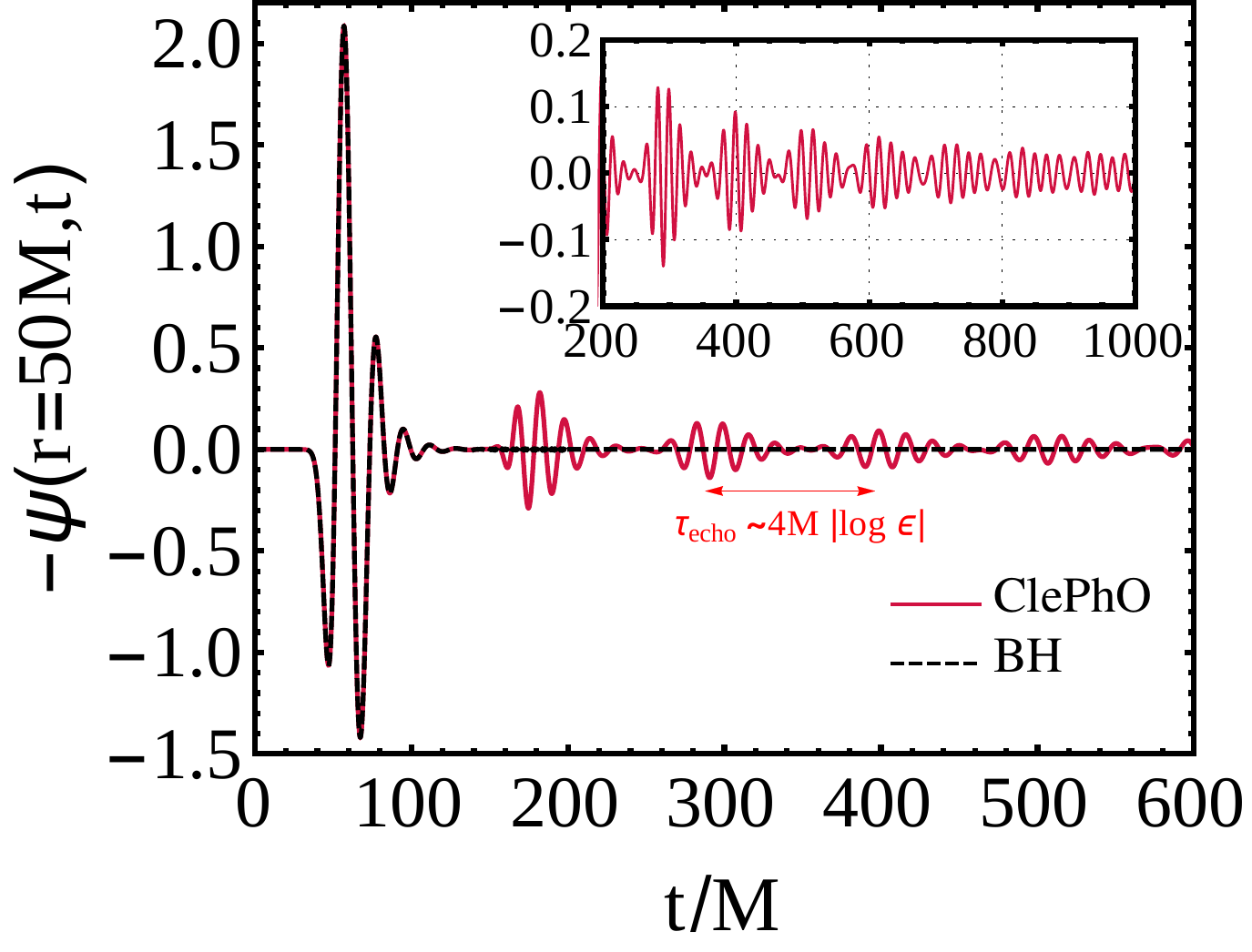}
\caption{Ringdown waveform for a BH (dashed black curve) compared to a ClePhO (solid red curve) with a reflective surface at $r_0=2M(1+\epsilon)$ with $\epsilon=10^{-11}$. We considered $l=2$ axial gravitational perturbations and a Gaussian wavepacket $\psi(r,0)=0$, $\dot\psi(r,0)=e^{-(z-z_m)^2/\sigma^2}$ (with $z_m=9M$ and $\sigma=6M$) as initial condition. 
Note that each subsequent echo has a smaller frequency content and that the damping of subsequent echoes is much larger than the late-time QNM prediction ($e^{-\omega_I t}$ with $\omega_I M\sim 4\times 10^{-10}$ for these parameters).
Data available online~\cite{rdweb}.
\label{fig:ringdown}}
\end{center}
\end{figure*}

For a BH, it becomes clear that the excitation of the spacetime modes happens at the photosphere~\cite{Ferrari:1984zz}. The vibrations excited there travel outwards to possible observers or
down the event horizon. The structure of GW signals at late times is therefore expected to be relatively simple. This is shown in Fig.~\ref{fig:ringdown}, which refers to the scattering of a Gaussian pulse of axial quadrupolar modes off a BH. The pulse crosses the photosphere, and excites its modes. The ringdown signal, a fraction of which travels to outside observers, is to a very good level described by the lowest QNMs. The other fraction of the signal generated at the photosphere travels downwards and into the horizon. It dies off and has no effect on observables at large distances.

Contrast the previous description with the dynamical response of ultracompact objects for which condition~\eqref{eps_crit} is satisfied (i.e., a ClePhO) [cf.\ Fig.~\ref{fig:ringdown}]. The initial description of the photosphere modes still holds, by causality. Thus, up to timescales of the order $|z_0|\sim-M\log\epsilon$ (the roundtrip time of radiation between the photosphere and $r_0$) the signal is {\it identical} to that of BHs~\cite{Cardoso:2016rao,Cardoso:2016oxy}.
At later times, however, the pulse traveling inwards is bound to see the object and be reflected either at its surface or at its center. In fact, this pulse is semi-trapped between the object and the light ring. Upon each interaction with the light ring, a fraction exits to outside observers, giving rise to a series of {\it echoes} of ever-decreasing amplitude. From Eqs.~\eqref{omegaRecho}--\eqref{omegaIecho}, repeated reflections occur in a characteristic echo delay time~\cite{Cardoso:2016rao,Cardoso:2016oxy}
\begin{equation}
 \tau_{\rm echo}\sim 4M |\log\epsilon|\,. \label{tauecho}
\end{equation}
However, the main burst is typically generated at the photosphere and has therefore a \emph{frequency content} of the same order as the BH QNMs~\eqref{bh_qnm}.
The initial signal is of high frequency and a substantial component is able to cross the potential barrier. Thus, asymptotic observers see a series of echoes whose amplitude is getting smaller and whose frequency content is also going down~\footnote{
An interesting, nontrivial example was worked out years ago in a different context~\cite{Ferrari:2000sr}. The example concerns axial GWs emitted during the scatter of point particles around ultracompact, constant-density stars. The GW signal shows a series of visible, well-space echoes after the main burst of radiation, which are associated to the quasi-trapped $s$-modes of ultracompact stars~\cite{Chandrasekhar449}. It turns out that all the results of that work fit very well our description: axial modes do not couple to the fluid and travel free to the geometrical center of the star, which is therefore the effective surface in this particular case. The time delay of the echoes in Fig.~1 of Ref.~\cite{Ferrari:2000sr} is very well described by the GW's roundtrip time to the center, $\tau_{\rm echo}\sim \frac{27\pi}{8}\epsilon^{-1/2}M$, where $r_0=\frac{9}{4}M(1+\epsilon)$ is the radius of the star and $r_0=\frac{9}{4}M$ is the Buchdahl's limit~\cite{Buchdahl:1959zz}.
}. At {\it very late times}, the QNM analysis is valid, and the signal is described by Eqs.~\eqref{omegaRecho}--\eqref{omegaIecho}. 

A formal description of the the frequency content of the sequence of echoes, and how they can be obtained with the help of the BH response, can be found in Ref.~\cite{Mark:2017dnq}.
A method to reconstruct the potential of the echo source, after the modes have been extracted was proposed in Ref.~\cite{Volkel:2017kfj}.

% 

%%%%%%%%%%%%%%%%%%%%%%%%%%%%%%%%%%%%%%%%%%%%%%%%%%%%%%%%%%%%%%%%%%%%%%%%%%%%%%
\subsection{The role of the spin}
%%%%%%%%%%%%%%%%%%%%%%%%%%%%%%%%%%%%%%%%%%%%%%%%%%%%%%%%%%%%%%%%%%%%%%%%%%%%%%
When spherical symmetry is broken by angular momentum $J$, Birkhoff's theorem does not apply, so the exterior metric of a spinning object is not unique and technically much more difficult to compute.\footnote{Nonetheless, there are indications that all multipole moments of the external spacetime approach those of a Kerr BH as $\epsilon\to0$~\cite{Pani:2015tga,Yagi:2015cda}. In this limit, it is natural to expect that the exterior geometry is extremely close to Kerr, unless some discontinuity occurs in the BH limit. It would be extremely important to confirm or confute this conjecture.}
Furthermore, BH uniqueness theorems do not apply in the presence of matter. In particular, a stationary object is not necessarily axisymmetric, unlike the Kerr solution which uniquely describes a BH in stationary configurations.
Even restricting to axisymmetric configurations, the linear response of a spinning object
depends also on the azimuthal number $m$ (such that $|m|\leq l$), which characterizes the angular dependence of the perturbations and yields a Zeeman-like splitting of the QNM frequencies. The relation between null geodesics and BH QNMs in the eikonal limit is more involved but conceptually similar to the static case~\cite{Yang:2012he}.

Assuming that the exterior spacetime approaches the Kerr geometry when $\epsilon\to0$, the linear response of a ClePhO can be studied with a toy model~\cite{Cardoso:2008kj,Abedi:2016hgu,Maggio:2017ivp,Nakano:2017fvh} in which the Kerr horizon is replaced by a reflective surface at $r_0=r_+(1+\epsilon)$, where $r_\pm=M(1\pm\sqrt{1-\chi^2})$ are the locations of the horizons and $\chi=J/M^2$ is the dimensionless spin. 
In this case, Eq.~\eqref{omegaRecho} is modified by the replacement $\omega_R\to\omega_R-m\Omega$ on the left-hand side~\cite{Vilenkin:1978uc} (where $\Omega$ is the angular velocity of the object when $\epsilon\to0$), whereas Eq.~\eqref{omegaIecho} is replaced by~\cite{Starobinskij2} 
\begin{equation}
M\omega_I \simeq -\frac{\beta_{ls}}{|z_0|}\left(\frac{2M^2r_+}{r_+-r_-}\right)\left[\omega_R(r_+-r_-)\right]^{2l+1}(\omega_R-m\Omega)\,, \label{omegaIechospin}
\end{equation}
where now $z_0\sim M[1+(1-\chi^2)^{-1/2}]\log\epsilon$. The result \eqref{omegaIechospin} shows how angular momentum can bring about substantial qualitative changes. The spacetime is unstable for $\omega_R(\omega_R-m\Omega)<0$ (i.e., in the superradiant regime~\cite{Brito:2015oca}), on a timescale $\tau_{\rm inst}\equiv 1/\omega_I$. This phenomenon is called \emph{ergoregion instability}~\cite{Friedman:1978wla,Brito:2015oca,Moschidis:2016zjy}.
In the $\epsilon\to0$ limit and for sufficiently large spin, $\omega_R\sim m\Omega$ and $\omega_I\sim |\log\epsilon|^{-1}$.
Notice that such description is only valid at small frequencies, and
therefore becomes increasingly less accurate at large spins and away
from the instability threshold (where $\omega_R,\omega_I\sim0$). For
large spin the result is more complex and can be found in Ref.~\cite{Hod:2017cga}.
Furthermore,  in the superradiant regime the ``damping'' factor, $\omega_I/\omega_R>0$, so that, at very late times (when the pulse frequency content is indeed described by these formulas), the amplitude of the QNMs \emph{increases} due to the instability. Such increase is anyway small, for example $\frac{\omega_I}{\omega_R}\approx 4\times10^{-6}$ when $\epsilon=0.001$, $l=2$ and $\chi=0.7$.

In the spinning case, the echo delay time~\eqref{tauecho} reads
%%%
\begin{equation}
 \tau_{\rm echo}\sim 2M[1+(1-\chi^2)^{-1/2}]\log\epsilon\,, \label{tauechospin}
\end{equation}
%%%
which corresponds to the period of the \emph{corotating} mode, $\tau_{\rm echo}\sim (\omega_R-m\Omega)^{-1}$.

Note that, as we explained earlier, the signal can only be considered as a series of well-defined pulses at early stages.
In this stage, the pulse still contains a substantial amount of high-frequency components which are not superradiantly amplified.
Thus, amplification will occur at late times; the early-time evolution of the pulse generated at the photosphere is more complex.
%%%%%%%%%%%%%%%%%%%%%%%%%%%%%%%%%%%%%%%%%%%%%%%%%%%%%%%%%%%%%%%%%%%%%%%%%%%%%%
\subsection{BHs in binaries: tidal heating and tidal deformability \label{Sec:tides}}
%%%%%%%%%%%%%%%%%%%%%%%%%%%%%%%%%%%%%%%%%%%%%%%%%%%%%%%%%%%%%%%%%%%%%%%%%%%%%%
The properties of an event horizon have also important consequences for the dynamics of binary systems containing a BH. 
As previously discussed, a spinning BH absorbs radiation of frequency $\omega>m\Omega$, but amplifies radiation of smaller frequency.
In this respect, BHs are dissipative systems which behave just like a Newtonian viscous fluid~\cite{Poisson:2009di,Cardoso:2012zn}.
Dissipation gives rise to various interesting effects in a binary system --~such as tidal heating~\cite{Hartle:1973zz,PhysRevD.64.064004}, tidal acceleration, and tidal locking~-- the Earth-Moon system being no exception due to the friction of the oceans with the crust.

For low-frequency circular binaries, the energy flux associated to tidal heating at the horizon corresponds to the rate of change of the BH mass~\cite{Alvi:2001mx,Poisson:2004cw},
\begin{eqnarray}
 \dot M = \dot E_H \propto \frac{\Omega_K^5}{M^2}(\Omega_K-\Omega)\,, \label{Edot}
\end{eqnarray}
where $\Omega_K\ll 1/M$ is the orbital angular velocity and the (positive) prefactor depends on the masses and spins of the two bodies. Thus, tidal heating is stronger for highly spinning bodies relative to the nonspinning by a factor $\sim \Omega/\Omega_K\gg1$.

The energy flux~\eqref{Edot} leads to a potentially observable phase shift of GWs emitted during the inspiral. Thus, it might be argued that an ECO binary can be distinguished from a BH binary, because $\dot E_H=0$ for the former~\cite{Maselli:2017cmm}. However, the trapping of radiation in ClePhOs can efficiently mimic the effect of a horizon~\cite{Maselli:2017cmm}. In order for absorption to affect the orbital motion, it is necessary that the time radiation takes to reach the companion, $T_{\rm rad}$, be much longer than the radiation-reaction time scale due to heating, $T_{\rm RR}\simeq E/\dot E_H$, where $E\simeq -\frac{1}{2} M(M\Omega_K)^{2/3}$ is the binding energy of the binary (assuming equal masses).
For BHs, $T_{\rm rad}\to\infty$ because of time dilation, so that the condition $T_{\rm rad}\gg T_{\rm RR}$ is always satisfied. 
For ClePhOs, $T_{\rm rad}$ is of the order of the GW echo delay time, Eq.~\eqref{tauecho}, and therefore increases logarithmically as $\epsilon\to0$. Thus, an effective tidal heating might occur even in the absence of an event horizon if the object is sufficient compact. The critical value of $\epsilon$ increases strongly as a function of the spin. For orbital radii larger than the ISCO, the condition $T_{\rm rad}\gg T_{\rm RR}$ requires $\epsilon\ll10^{-88}$ for $\chi\lesssim0.8$, and therefore even Planck corrections at the horizon scale are not sufficient to mimic tidal heating. This is not necessarily true for highly spinning objects, for example $T_{\rm rad}\gg T_{\rm RR}$ at the ISCO requires $\epsilon\ll10^{-16}$ for $\chi\approx0.9$.

%%%

Finally, the nature of the inspiralling objects is also encoded in the way they respond when acted upon by the external gravitational field of their companion~-- through their tidal Love numbers~\cite{PoissonWill}. An intriguing result in classical GR is the fact that the tidal Love numbers of a BH are precisely zero~\cite{Binnington:2009bb,Damour:2009vw,Porto:2016zng}. On the other hand, those of ECOs are small but finite~\cite{Cardoso:2017cfl}. In particular, the tidal Love numbers of ClePhOs vanish logarithmically in the BH limit. Thus, any measurement of the tidal Love number $k$ translates into an estimate of the distance of the ECO surface from its Schwarzschild radius,
%%%
\begin{equation}
 \epsilon\sim e^{-1/k}\,.
\end{equation}
Owing to the above exponential dependence, a measurement of the tidal Love number of an ECO at the level of $k\approx{\cal O}(10^{-3})$ already probes Planck distances away from the gravitational radius $r_g$~\cite{Cardoso:2017cfl}. This level of accuracy is in principle within reach future GW detectors~\cite{Maselli:2017cmm}.

\newpage
%%%%%%%%%%%%%%%%%%%%%%%%%%%%%%%%%%%%%%%%%%%%%%%%%%%%%%%%%%%%%%%%%%%%%%%%%%%%%%
\section{Beyond vacuum black holes}
%%%%%%%%%%%%%%%%%%%%%%%%%%%%%%%%%%%%%%%%%%%%%%%%%%%%%%%%%%%%%%%%%%%%%%%%%%%%%%

\hskip 0.2\textwidth
\parbox{0.8\textwidth}{
\begin{flushright}
{\small 
\noindent {\it ``Mumbo Jumbo is a noun and is the name of a grotesque idol said to have been worshipped by some tribes. In its figurative sense, Mumbo Jumbo is an object of senseless veneration or a meaningless ritual.''}\\
Concise Oxford English Dictionary
}
\end{flushright}
}

\vskip 1cm

To entertain the possibility that dark, massive, compact objects are not BHs, requires
one to discuss some outstanding issues. One can take two different stands on this topic:

\begin{itemize}
 \item[i.] a pragmatic approach of testing the spacetime close to compact, dark objects, irrespective
of their nature, by devising model-independent observations that yield unambiguous answers.
%%%
 \item[ii.] a less ambitious and more theoretically-driven approach, which starts by constructing objects that are very compact, yet horizonless, within some framework. It proceeds to study their formation mechanisms and stability properties; then discard solutions which either do not form or are unstable on relatively short timescales; finally, understand the observational imprints of the remaining objects, and how they differ from BHs'.
\end{itemize}

In practice, when dealing with outstanding problems where our ignorance is extreme, both approaches should be used simultaneously.
Indeed, using concrete models can sometimes be a useful guide to learn about broad, model-independent signatures.
As it will hopefully become clear, one could design exotic horizonless models which mimic all observational properties of a BH with arbitrary accuracy. 
While the statement ``BHs exist in our Universe'' is \emph{fundamentally unfalsifiable}\footnote{Arguably, this is true for any statement about the \emph{existence} of a physical entity in our Universe, but this does not prevent us to accumulate more and more evidence to support a given model or theory, nor to rule out competitors.}, alternatives can be ruled out or confirmed to exist with a single observation, just like Popper's black swans.

%%%%%%%%%%%%%%%%%%%%%%%%%%%%%%%%%%%%%%%%%%%%%%%%%%%%%%%%%%%%%%%%%%%%%%%%%%%%%%
\subsection{Are there alternatives?}
%%%%%%%%%%%%%%%%%%%%%%%%%%%%%%%%%%%%%%%%%%%%%%%%%%%%%%%%%%%%%%%%%%%%%%%%%%%%%%

%
\begin{table}[ht!]
%\capstart
\begin{small}
\begin{tabular}{@{}@{}l@{}|@{}c@{}@{}c@{}@{}c@{}@{}c@{}@{}c@{}@{}}
\hline \noalign{\smallskip}\hline \noalign{\smallskip}
%
%\multicolumn{6}{c}{${\tilde{\omega}}$}\\ 
Model   & Taxonomy & Formation  & Stability & EM signatures & GWs \\ 
\noalign{\smallskip}
\hline 
%
%Neutron stars    & CO                                                                                 &\yes &\yes  &\yes\\
%                 &                                                                                    & &   & \\
%
\\
Fluid stars      & UCOs                                        &\no                           &\yes                                                                                     &\yes & \yes\\
                 &                                           &                              & \cite{Kokkotas:1999bd,Cardoso:2014sna,Saida:2015zva,Stuchlik:2017qiz,Volkel:2017ofl,Volkel:2017kfj}   &   &\cite{Kokkotas:1999bd,Ferrari:2000sr,Cardoso:2014sna,Volkel:2017ofl} \\						
\\
Anisotropic stars      & ClePhOs                                        &\no                           &\yes                                                                                     &\yes & \yes\\
	               & \cite{1974ApJ...188..657B,Dev:2000gt,Silva:2014fca}                           &                              & \cite{Dev:2003qd,Doneva:2012rd}   &\cite{Silva:2014fca,Yagi:2015upa,Yagi:2015cda}   &\cite{Yagi:2015upa,Yagi:2015cda} \\						
\\
Boson stars \&   & UCOs, (ClePhOs?)                             &\yes                          &\yes                                           & \yes &\yes \\
oscillatons      & ~\cite{Kaup:1968zz,Ruffini:1969qy,Colpi:1986ye,Seidel:1991zh,Brito:2015yga,Brito:2015pxa,Liebling:2012fv,Grandclement:2016eng}  &\cite{Seidel:1991zh,Seidel:1993zk,Okawa:2013jba,Brito:2015yfh,Liebling:2012fv} & \cite{Gleiser:1988ih,Lee:1988av,Honda:2001xg,Cardoso:2007az,Brito:2015pxa,Macedo:2013jja}   &\cite{Vincent:2015xta,Cao:2016zbh,Shen:2016acv}  &\cite{Palenzuela:2007dm,Kesden:2004qx,Choptuik:2009ww,Macedo:2013qea,Cardoso:2016oxy,Cardoso:2017cfl,Sennett:2017etc,Maselli:2017cmm}\\
\\
Gravastars       & COs -- ClePhOs                                  &\no                             & \yes                                           &\yes & $\sim$\\
                 & \cite{Mazur:2001fv,Mazur:2004fk}         &                                &\cite{Cardoso:2007az}                          & \cite{Sakai:2014pga,Uchikata:2015yma,Uchikata:2016qku} & \cite{Chirenti:2007mk,Pani:2009hk,Pani:2009ss,Pani:2010em,Chirenti:2016hzd,Cardoso:2016rao,Cardoso:2016oxy,Uchikata:2016qku,Cardoso:2017cfl,Maselli:2017cmm,Volkel:2017ofl,Volkel:2017kfj}  \\
\\
AdS bubbles      & UCOs -- ClePhOs                           &\no                             & \yes                                           &$\sim$ & \no\\
                 & \cite{Danielsson:2017riq}         &                                &\cite{Danielsson:2017riq}                          & \cite{Danielsson:2017riq} &   \\
\\
Wormholes       & ClePhOs                                   &\no                                                                              & \yes                                  & \yes & $\sim$\\
                 & \cite{Morris:1988tu,VisserBook,Lemos:2003jb,Damour:2007ap,Lemos:2008cv} &             &\cite{Gonzalez:2008wd,Gonzalez:2008xk}         &\cite{Nedkova:2013msa,Ohgami:2015nra,Abdujabbarov:2016efm,Zhou:2016koy}& \cite{Cardoso:2016rao,Cardoso:2017cfl,Maselli:2017cmm}  \\
\\
Fuzzballs        & ClePhOs                                  & \no 	                         &	\no                                               & \no                                &$\sim$\\
                 & \cite{Mathur:2005zp,Mathur:2008nj}       &                                & (but see \cite{Cardoso:2005gj,Chowdhury:2007jx,Eperon:2016cdd,Eperon:2017bwq})   &                                    &(but see 
								\cite{Cardoso:2016rao,Cardoso:2016oxy,Hertog:2017vod})\\
\\
Superspinars     & COs -- ClePhOs                           &\no	                             &\yes	                                              & \no                         & $\sim$\\
                 & \cite{Gimon:2007ur}                     &                                 & \cite{Cardoso:2008kj,Pani:2010jz}                  & (but see \cite{Patil:2015fua})&  \cite{Cardoso:2016rao,Cardoso:2016oxy} \\
\\
$2-2$ holes      & ClePhOs 	                                & \no	                           & \no	                                               &\no                                & $\sim$\\
~                & \cite{Holdom:2016nek}                    &                                & (but see \cite{Holdom:2016nek})                     & (but see \cite{Holdom:2016nek})   & \cite{Cardoso:2016rao,Cardoso:2016oxy}\\								
\\
Collapsed   & ClePhOs 	                                & \no	                         & \no	                                 &\no                      & $\sim$\\
polymers    & \cite{Brustein:2016msz,Brustein:2017kcj}                    		 &                               & (but see \cite{Brustein:2016msz,Brustein:2017koc})                      			 & 				     & ~\cite{Brustein:2017koc}\\								
\\
Quantum bounces /  & ECO -- ClePhOs 	                                & \no	                         & \no	                                 &\no                                & $\sim$\\
black stars & \cite{Malafarina:2017csn,Barcelo:2007yk,Barcelo:2009tpa,Barcelo:2015noa,Barcelo:2017lnx,Kawai:2017txu}                     & (but see~\cite{Bambi:2013caa,Barcelo:2007yk})                               &                 & 	& ~\cite{Barcelo:2017lnx}\\								
\\
Quantum stars$^{*}$ & UCOs -- ClePhOs                                 &\no	                            &\no	                                               &\no                             &\no\\
~                & \cite{Rovelli:2014cta,Baccetti:2017ioi}                                      &                                &                                                     &                                 &  \\
Fire-walls$^{*}$ & ClePhOs	                                 &\no	                           &\no	                                                &\no                                  &$\sim$ \\
~                & \cite{Almheiri:2012rt,Zurek:1984zz,Braunstein:2009my}                    &                               &                                                    &                                     &\cite{Barausse:2014tra,Cardoso:2016oxy}  \\
\noalign{\smallskip}\hline \noalign{\smallskip} \hline
\end{tabular}
\end{small}
\vskip 8pt 
% \scriptsize
\centering \caption{Catalogue of some proposed horizonless compact objects. A $\yes$ tick means that the topic was addressed. With the exception of boson stars, however,
most of the properties are not fully understood yet. An asterisk $^*$ stands for the fact that these objects {\it are} BHs, but could have phenomenology similar to
the other compact objects in the list. 
This table does not include models of quantum-corrected BHs (e.g.,~\cite{Dvali:2011aa,Dvali:2012rt,Giddings:2014ova}), even when the latter predict large corrections near the horizon, because also in this case the object is a BH (although different from Kerr).
}
\label{tab:ECOs}
\end{table}

%}
A nonexhaustive summary of possible objects which could mimic BHs is shown in Table~\ref{tab:ECOs}. Stars made of constant density fluids
are perhaps the first known example of compact configurations, the literature on the subject is too long to list.
In GR, isotropic static spheres made of ordinary fluid satisfy the Buchdahl limit on their compactness, $2M/r_0<8/9$~\cite{Buchdahl:1959zz}, and can thus never be a ClePhO.
Interestingly, polytrope UCO stars {\it always} have superluminal sound speed~\cite{Saida:2015zva}.

Compact solutions can be built with fundamental, massive bosonic fields. The quest for self-gravitating 
bosonic configurations started in an attempt to understand if gravity could 
produce ``solitons'' through the nonlinearities of the field equations. Such configurations are broadly referred to as boson stars. Boson stars have attracted considerable interest
as light scalars are predicted to occur in different scenarios, and ultralight scalars can explain the dark matter puzzle~\cite{Hui:2016ltb}.
When fastly spinning, such stars can be extremely compact. There seem to be no studies on the classification of such configurations (there are solutions known to display
photospheres, but it is unknown whether they can be as compact as ClePhOs).

The other objects in the list require either unknown matter or large quantum effects. For example,
a specific model of a wormhole was discussed carefully
in the context of ``BH foils''~\cite{Damour:2007ap}. In the nonspinning case, the geometry of this model is described by 
\begin{equation}
ds^2=-\left(f+\lambda\right)dt^2+\frac{dr^2}{f}+r^2d\Omega^2\,.
\end{equation}
The constant $\lambda$ is assumed to be extremely small, $\lambda\sim e^{-M^2/\ell_P^2}$ where $\ell_P$ is the Planck length. In the context of GR, this solution requires matter violating the null energy condition.
Other objects, such as fuzzballs, were introduced as a microstate description of BHs in string theory. In this setup the individual microstates
are horizonless and the horizon arises as a coarse-grained description of the microstate geometries. 
Phenomena such as Hawking radiation can be recovered, in some instances,
from classical instabilities~\cite{Chowdhury:2007jx,Chowdhury:2008uj}.
``Gravitational-vacuum stars'' or {\it gravastars}~\cite{Mazur:2001fv,Mazur:2004fk} are ultracompact configurations supported by a negative pressure, which might arise 
as an hydrodynamical description of one-loop QFT effects in curved spacetime, so they do not necessarily require exotic new physics~\cite{Mottola:2006ew}. In these models, the Buchdahl limit is evaded because the internal effective fluid is anisotropic and also violates the energy conditions~\cite{Mazur:2015kia}. Gravastars have been recently generalized to include anti-de Sitter cores,
in what was termed {\it AdS bubbles}, and which may allow for holographic descriptions~\cite{Danielsson:2017riq}.

For most objects listed in Table~\ref{tab:ECOs} which are inspired by quantum-gravity corrections, the changes in the geometry occur
deep down in the strong field regime. Some of these models --~including also black stars~\cite{Barcelo:2009tpa}, superspinars~\cite{Gimon:2007ur} or collapsed polymers~\cite{Brustein:2016msz,Brustein:2017kcj}~-- predict that large quantum backreaction should affect the horizon geometry even for macroscopic objects. In these models, our parameter $\epsilon$ is naturally of the order $\sim{\cal O}(\ell_P/M)\in(10^{-39},10^{-46})$ for masses in the range $M\in(10,10^8)M_\odot$.
The $2-2$-hole model predicts even more compact objects, with $\epsilon\sim (\ell_P/M)^2\in(10^{-78},10^{-92})$~\cite{Holdom:2016nek}.
In all these cases, both quantum-gravity or microscopic corrections at the horizon scale select ClePhOs as well-motivated alternatives to BHs.

Despite a number of supporting arguments --~some of which urgent and well founded~-- it is important to highlight that there is no horizonless ClePhO for which we know sufficiently well the physics at the moment.

%%%%%%%%%%%%%%%%%%%%%%%%%%%%%%%%%%%%%%%%%%%%%%%%%%%%%%%%%%%%%%%%%%%%%%%%%%%%%%
\subsection{Formation and evolution}
%%%%%%%%%%%%%%%%%%%%%%%%%%%%%%%%%%%%%%%%%%%%%%%%%%%%%%%%%%%%%%%%%%%%%%%%%%%%%%
Although supported by sound arguments, the vast majority of the alternatives to BHs are, at best, incompletely described. Precise calculations (and often even a rigorous framework) incorporating the necessary physics are missing.

One notable exception to our ignorance are boson stars. These configurations are known to arise, generically, out of the gravitational
collapse of massive scalars. Their interaction and mergers can be studied by evolving the Einstein-Klein-Gordon system, and there is evidence that accretion of less massive boson stars
makes them grow and cluster around the configuration of maximum mass. In fact, boson stars have efficient {\it gravitational cooling}
mechanisms that allow them to avoid collapse to BHs and remain very compact after interactions~\cite{Seidel:1991zh,Seidel:1993zk,Brito:2015yfh}.

The remaining objects listed in Table~\ref{tab:ECOs} were built in a phenomenological way or they arise as solutions of Einstein equations coupled to exotic matter fields. For example, models of quantum-corrected objects do not include all the (supposedly large) local or nonlocal quantum effects
that might prevent collapse from occurring. In the absence of a complete knowledge of the missing physics, it is unlikely that a ClePhO forms out of the merger of two ClePhOs. These objects are so compact that at merger they will be probably engulfed by a common apparent horizon. The end product is, most likely, a BH.  
On the other hand, if large quantum effects do occur, they would probably act on short timescales
to prevent apparent horizon formation. In some models, Planck-scale dynamics naturally leads to abrupt changes close to the would-be horizon, without fine tuning~\cite{Holdom:2016nek}. Likewise, in the presence of (exotic) matter or if GR is classically modified at the horizon scale, Birkhoff's theorem no longer holds, and a star-like object might be a more natural outcome than a BH.
In summary, with the exception of boson stars, we do not know how or if UCOs and ClePhOs form in realistic collapse or merger scenarios.
\clearpage
\newpage
%%%%%%%%%%%%%%%%%%%%%%%%%%%%%%%%%%%%%%%%%%%%%%%%%%%%%%%%%%%%%%%%%%%%%%%%%%%%%%
\subsection{On the stability problem}
%%%%%%%%%%%%%%%%%%%%%%%%%%%%%%%%%%%%%%%%%%%%%%%%%%%%%%%%%%%%%%%%%%%%%%%%%%%%%%

\hskip 0.2\textwidth
\parbox{0.8\textwidth}{
\begin{flushright}
{\small 
\noindent {\it ``There is nothing stable in the world; uproar's your only music.''}\\
% \vskip 2mm
John Keats, Letter to George and Thomas Keats, Jan 13 (1818)
}
\end{flushright}
}

\vskip 1cm

Appealing solutions are only realistic if they form and remain as long-term stable solutions of the theory.
In other words, solutions have to be stable when slightly perturbed or they would not be observed. There are strong indications that the exterior Kerr spacetime is stable,
although a rigorous proof is still missing~\cite{Dafermos:2008en}.
There are good reasons to believe that some --~if not all~--
horizonless compact solutions are linearly or nonlinearly unstable. 

Some studies of linearized fluctuations of ultracompact objects are given in Table~\ref{tab:ECOs}. We will not discuss specific models, but we would like to highlight two general results. Linearized gravitational fluctuations of any nonspinning UCO are extremely long-lived and decay no faster than logarithmically~\cite{Keir:2014oka,Cardoso:2014sna,Eperon:2016cdd,Eperon:2017bwq}. Indeed, such perturbations can be again understood in terms of modes quasi-trapped within the potential barrier shown in Fig.~\ref{fig:potential}: they require a photosphere but are absent in the BH case. For a ClePhO, these modes are very well approximated by Eqs.~\eqref{omegaRecho}--\eqref{omegaIecho}.
The long damping time of these modes has led to the conjecture that any UCO is nonlinearly unstable and may evolve through a Dyson-Chandrasekhar-Fermi type of mechanism~\cite{Keir:2014oka,Cardoso:2014sna}.
The endstate is unknown, and most likely depends on the equation of the state of the particular UCO: some objects may fragment and evolve past the UCO region into less compact configurations,
via mass ejection, whereas other UCOs may be forced into gravitational collapse to BHs. 

The above mechanism is supposed to be active for any spherically symmetric UCO, and also on spinning solutions. However, it is nonlinear in nature. On the other hand, UCOs (and especially ClePhOs) can develop negative-energy regions once spinning. In such a case, there is a well-known {\it linear}
instability, dubbed as ergoregion instability. The latter affects any horizonless geometry with an ergoregion~\cite{Friedman:1978wla,Moschidis:2016zjy,Cardoso:2007az,Oliveira:2014oja,Maggio:2017ivp} and is deeply connected to superradiance~\cite{Brito:2015oca}.

In summary, there is good evidence that UCOs are linearly or nonlinearly unstable. Unfortunately, the effect of viscosity is practically unknown~\cite{Cardoso:2014sna},
and so are the timescales involved in putative dissipation mechanisms that might quench this instability. On the other hand, even unstable solutions are relevant if the timescale is larger than any other dynamical scale in the problem~\footnote{After all, ``we are all unstable''.}. The nonlinear mechanism at work for spherically symmetric spacetimes presumably acts on very long timescales only; a model problem predicts an exponential dependence on the size of the initial perturbation~\cite{FritzJohn}.
At least in some portion of the parameter space the ergoregion-instability timescale is very large~\cite{Friedman:1978wla,Cardoso:2007az,Maggio:2017ivp}.
%%%
From Eq.~\eqref{omegaIechospin}, we can estimate the timescale of the instability of a spinning ClePhO,
%%%
\begin{equation}
 \tau\equiv\frac{1}{\omega_I}\sim-|\log\epsilon|\frac{1+(1-\chi^2)^{-1/2}}{2\beta_{ls}}\left(\frac{r_+-r_-}{r_+ }\right) \frac{\left[\omega_R(r_+-r_-)\right]^{-(2l+1)}}{\omega_R-m\Omega}\,. \label{tauinstab}
\end{equation}
%%%%
As previously discussed, a spinning ClePhO is (superradiantly) unstable only when $\Omega>\omega_R/m$. For example, for $l=m=s=2$ and $\chi=0.7$, the above formula yields
%%%%
\begin{equation}
 \tau\in\left(5,1\right)\left(\frac{M}{10^6M_\odot}\right)\,{\rm yr}\quad {\rm when~~} \epsilon\in (10^{-45},10^{-22})\,.
\end{equation}
%%%%%
Generically, the ergoregion instability acts on timescales which are parametrically longer than the dynamical timescale, $\sim M$, of the object. Given such long timescales, it is likely that the instability can be efficiently quenched by some dissipation mechanism of nongravitational nature, although this effect would be model-dependent~\cite{Maggio:2017ivp}. Furthermore, it is possible that the endstate of the instability is simply an ECO spinning at the superradiant threshold, $\Omega=\omega_R/m$. In such case, the instability would only rule out highly-spinning ECOs in a certain compactness range. Likewise, EM or GW observations indicating statistical prevalence of slowly-spinning compact objects, across the entire mass range, could be an indication for the absence of an horizon.

Finally, there are indications that instabilities of UCOs are merely the equivalent of Hawking radiation for these geometries, and that therefore
there might be a smooth transition in the emission properties when approaching the BH limit~\cite{Chowdhury:2007jx,Damour:2007ap}.

\newpage

%%%%%%%%%%%%%%%%%%%%%%%%%%%%%%%%%%%%%%%%%%%%%%%%%%%%%%%%%%%%%%%%%%%%%%%%%%%%%%%%%%%%%%
\section{Have exotic compact objects been already ruled out by electromagnetic observations?}
%%%%%%%%%%%%%%%%%%%%%%%%%%%%%%%%%%%%%%%%%%%%%%%%%%%%%%%%%%%%%%%%%%%%%%%%%%%%%%%%%%%%%%
%

\hskip 0.2\textwidth
\parbox{0.8\textwidth}{
\begin{flushright}
{\small 
\noindent {\it ``And here - ah, now, this really is something a little recherch\'e.''}\\
% \vskip 2mm
Sherlock Holmes, {\it The Musgrave Ritual}, Sir Arthur Conan Doyle
}
\end{flushright}
}

\vskip 1cm

Before the GW revolution, questions about the observational evidence for BHs were initially posed in the context of EM observations. A concise overview of the arguments is presented in Refs.~\cite{Abramowicz:2002vt,2017FoPh..tmp...26E}, concluding that ``it is fundamentally impossible to give an observational proof for the existence of a BH horizon.'' Surprisingly however, there have been recent claims~\cite{Broderick:2005xa,Broderick:2007ek,Broderick:2009ph} that observations of some objects (most notably the compact radio source Sgr~A$^*$) all but rule out other candidates, {\it including} those whose surface is a Planck-distance away from the horizon (i.e., $\epsilon\sim 10^{-45}$ for Sgr~A$^*$, thus qualifying as ClePhOs according to our classification). 
The argument can, roughly, be described as follows:
%%%
\begin{itemize}
 \item[i.] There are systems with a very low accretion rate and very low luminosity, for which the central object is very dark. 
%This is a fact.
%%%
 \item[ii.] The coordinate time that a particle in the accretion disk takes to reach the surface scales as $T_{\rm travel}\sim M |\log \epsilon|$. Even for Planck-sized $\epsilon$ this time is very small. 
%This is also a fact.
%%%
 \item[iii.] Assumption: because the above timescale is so small, a thermodynamic and dynamic equilibrium must be established between the accretion disk and the central object, on relatively short timescales.
%%%
 \item[iv.] Assumption: Because there is equilibrium, the central object must be returning in EM radiation most of the energy that it is taking in from the disk. 
%%%
 \item[v.] Because we see no such radiation, the premise that there is a surface must be wrong.
\end{itemize}
%%%

The arguments above fail to take into account important physics:

\paragraph{Timescale to reach equilibrium.} Strong lensing can prevent equilibrium from being achieved around ultracompact objects in sufficiently short time. Consider matter in the accretion disk falling in, releasing scattered radiation on the surface of the object which is then observed by our detectors. Suppose, for the sake of the argument, that once hitting the surface, it is scattered isotropically.
Then, as discussed in Section~\ref{sec:escape}, only a fraction $\sim \epsilon$ is able to escape during the first interaction with the star, cf.\ Eq.~\eqref{solidangle}. The majority of the radiation, will fall back onto the surface after a time $t_{\rm roundtrip} \sim 9.3 M$ given by the average of Eq.~\eqref{troundtrip}~\footnote{One might wonder if the trapped radiation bouncing back and forth the surface of the object might not interact with the accretion disk. As we showed in Section~\ref{sec:escape}, this does not happen, as the motion of trapped photons is confined to within the photosphere.}. Suppose one injects, instantaneously,
an energy $\delta M$ onto the object. Then, after a time $T$, the energy emitted to infinity during $N=T/t_{\rm roundtrip}$ interactions reads
%%%
\begin{equation}
 \Delta E \sim  \left[1 - (1 - \epsilon)^N\right]{\delta M}\,.\label{delta E}
\end{equation}
%%%
Clearly, $\Delta E\to \delta M$ as $N\to\infty$ since all energy will eventually escape to infinity. However, $\Delta E \sim \epsilon N \delta M$ if $\epsilon N\ll1$, i.e. when
%%%
\begin{equation}
 \epsilon\ll 10^{-16}\left(\frac{M}{10^6 M_\odot}\right)\left(\frac{t_{\rm Hubble}}{T}\right)\,, \label{condequilibrium}
\end{equation}
%%%
where we have been extremely conservative in normalizing $T$ by the age of the Universe, $t_{\rm Hubble}\approx10^{10}\,{\rm yr}$.
The luminosity of ClePhOs satisfying the above property is roughly
\begin{equation}
\dot{E}\sim 10^{-17}\left(\frac{\epsilon}{10^{-16}}\right)\left(\frac{\delta M}{M}\right)\,.
\end{equation}
When $\epsilon\ll10^{-16}$, this quantity is minute and it is impossible to achieve equilibrium on any meaningful timescale. Indeed, any injected energy $\delta M$ is eventually radiated in a timescale longer than $t_{\rm Hubble}$ whenever Eq.~\eqref{condequilibrium} is satisfied. Thus, gravastars, fuzzballs, and other objects inspired by quantum effects at the horizon scale ($\epsilon\lesssim 10^{-45}$) are not ruled out as exotic supermassive objects.

\noindent {\bf Cascade into other channels.} Even if the object were returning all of the incoming radiation on a sufficiently short timescale, a sizeable fraction of this energy could be in channels other than EM.
Let us estimate the center-of-mass energy in collisions between matter being accreted and the compact object. The infalling matter is assumed to be on radial free fall with four-velocity $v_{(1)}^{\mu}=(E/f,-\sqrt{E^2-f},0,0)$.
For particles at the surface of the object $v_{(2)}^{\mu}=(\/\sqrt{f},0,0,0)$.
When the particles collide, their CM energy reads~\cite{Banados:2009pr},
\be
E_{\rm CM}=m_0\sqrt{2}\sqrt{1-g_{\mu\nu}v_{(1)}^\mu v_{(2)}^\mu}\sim \frac{m_0\sqrt{2E}}{\epsilon^{1/4}}\,,
\ee
with $m_0$ the mass of the colliding particles (assumed to be equal). Therefore, the Lorentz factor of the colliding particles, in the CM frame, is $\gamma \gg 3\times 10^{5}$ for $\epsilon\ll10^{-22}$, and $\gamma\sim10^{11}$ for Planck-scale modifications of supermassive objects. Therefore, $E_{\rm CM}$ is orders of magnitude above the maximum energy achievable at the LHC.
At these CM energies, it is possible that new degrees of freedom are excited (some of which might even be the unknown physics that forms these compact objects in the first place). It is well possible that if new fields exist they carry a substantial fraction of the luminosity (while going undetected in our telescopes)\footnote{Incidentally, gravitational radiation makes only negligible contribution to the total flux, even for such large CM energies.
The ratio of EM to gravitational radiation generated during these processes was estimated to be~\cite{Cardoso:2003cn}
\be
\frac{E_{\rm EM}}{E_{\rm grav}}\sim \frac{\log\gamma}{\gamma}\frac{q}{m_0}\,,
\ee
with $q/m_0$ the charge-to-mass ratio of the colliding particles. Thus, even at such large CM energies, EM radiation is vastly larger than gravitational radiation, because $q/m_0\sim 10^{19}$ for protons.
}.
Even without advocating new physics beyond the $10\,{\rm TeV}$ scale, extrapolation of known hadronic interactions to large energies suggests that about $20\%$ of the collision energy goes into neutrinos, whose total energy is a sizeable fraction of that of the photons emitted in the process~\cite{Kelner:2006tc}.

In summary, we believe that some observations of supermassive objects most likely are compatible with ClePhOs (in particular they do not rule out gravastars nor quantum corrections at the horizon scale); these same observations are probably not compatible with the remaining of the UCO parameter space.

No argument whatsoever can be made to rule out exotic compact objects of \emph{arbitrary} compactness, simply because the transition to the BH limit should be continuous.
Therefore, \emph{any} argument suggesting that exotic compact objects are excluded for any $\epsilon$ has to be taken with great care.

\newpage

%%%%%%%%%%%%%%%%%%%%%%%%%%%%%%%%%%%%%%%%%%%%%%%%%%%%%%%%%%%%%%%%%%%%%%%%%%%%%%
\section{Testing the nature of BHs with GWs}\label{sec:GWs}
%%%%%%%%%%%%%%%%%%%%%%%%%%%%%%%%%%%%%%%%%%%%%%%%%%%%%%%%%%%%%%%%%%%%%%%%%%%%%%
%

\hskip 0.2\textwidth
\parbox{0.8\textwidth}{
{\small 
\noindent {\it ``It is well known that the Kerr solution provides the unique solution for stationary BHs in the universe.
But a confirmation of the metric of the Kerr spacetime (or some aspect of it) cannot even be contemplated in the foreseeable future.''}
% \vskip 2mm
\begin{flushright}
S. Chandrasekhar, The Karl Schwarzschild Lecture,\\ Astronomische Gesellschaft, Hamburg (September 18, 1986)
\end{flushright}
}

}

\vskip 1cm

Testing the nature of dark, compact objects with EM observations is plagued with several difficulties. Some of these are tied
to the incoherent nature of the EM radiation in astrophysics, and the amount of modeling and uncertainties associated to such emission. Other problems are connected to the absorption by
the interstellar medium. As discussed in the previous section, testing quantum or microscopic corrections at the horizon scale with EM probes is very challenging.

The historical detection of GWs by aLIGO~\cite{Abbott:2016blz} opens up the exciting possibility of testing gravity in extreme regimes with unprecedented accuracy~\cite{TheLIGOScientific:2016src,Yunes:2013dva,Barausse:2014tra,Berti:2015itd,Yunes:2016jcc,Maselli:2017cmm}. GWs are generated by coherent motion of massive sources, and
are therefore subjected to less modeling uncertainties (they depend on far fewer parameters) relative to EM probes. The most luminous GWs come from very dense sources, but they also interact very feebly with matter, thus providing the cleanest picture of the cosmos, complementary to that given by telescopes and particle detectors.

Compact binaries are the preferred sources for GW detectors (and in fact were the source of the first detected events~\cite{Abbott:2016blz,Abbott:2016nmj}). The GW signal from compact binaries is naturally divided
in three stages, corresponding to the different cycles in the evolution driven by GW emission~\cite{Buonanno:2006ui,Berti:2007fi,Sperhake:2011xk}:
the inspiral stage, corresponding to large separations and well approximated by post-Newtonian theory; the merger phase when the two objects coalesce and which can only be described accurately through numerical simulations; and finally, the ringdown phase when the
merger end-product relaxes to a stationary, equilibrium solution of the field equations~\cite{Sperhake:2011xk,Berti:2009kk,Blanchet:2013haa}.

All three stages provide independent, unique tests of gravity and of compact GW sources.
The GW signal in the two events reported so far is consistent with them being generated by BH binaries,
and with the endstate being a BH. To which level, and how, are alternatives consistent with current and future observations?
This question was recently addressed by several authors and may be divided into two different schemes (see Table~\ref{tab:summaryGW}).

\begin{table}[ht]
\centering
\begin{footnotesize}
\begin{tabular}{|c|c|c|c|c|}
\hline
\hline
&			        & BH 				& ECO 							    & ClePhO 					        \\
\hline
\multirow{4}{*}{\rotatebox[origin=c]{90}{\textbf{~ringdown~}}} & 			& 				& 					    & 		  \\
% & 			& 				& 					    & 		  \\
& GW echoes			& \no				& \yes (only UCOs)					    & \yes ($\tau_{\rm echo}\sim M|\log\epsilon|$)		  \\
& Modified prompt ringdown			& \no				& \yes					    & \no		  \\	
& Extra modes		& 	\no			& 	\yes				    & 	\yes	  \\
& 			& 				& 					    & 		  \\
\hline
\multirow{4}{*}{\rotatebox[origin=c]{90}{\textbf{inspiral}}}& Multipolar structure ($2$PN) 		& $\delta M_l=\delta S_l =0$    & $\delta M_l\neq0$, $\delta S_l \neq0$     		    & $\delta M_l\simeq0$, $\delta S_l \simeq0$              \\
& Tidal heating ($2.5-4$PN)        		& \yes			        & \no          					            & \no   					         \\
& Tidal Love number	($5$PN)                & $k=0$				& $k\lesssim{ \cal O}(k_{\rm NS})$          		    & $k\sim[\log\epsilon]^{-1}$		             \\ 
& Resonances	                & \no				& \cite{Macedo:2013jja,Pani:2010em}         & $\omega M\sim[\log\epsilon]^{-1}$	                \\ 
\hline
\hline
\end{tabular}
\caption{Summary of the properties of exotic compact objects that affect the GW signal relative to BHs in the ringdown phase and in the inspiral phase at a given post-Newtonian (PN) order. In the multipolar structure, $\delta M_l$ and $\delta S_l$ are the corrections to the mass and current multipole moments relative to the no-hair relation for a Kerr BH in isolation~\cite{Hansen:1974zz}, $M_l+i S_l =M^{l+1}\left(i\chi\right)^l$. 
Tidal absorption at the horizon enters at $2.5$PN ($4$PN) order for spinning (nonspinning) BHs.
The typical tidal Love numbers of a neutron star are denoted by $k_{\rm NS}$.}\label{tab:summaryGW}
\end{footnotesize}
\end{table}

%%%%%%%%%%%%%%%%%%%%%%%%%%%%%%%%%%%%%%%%%%%%%%%%%%%%
\subsection{Smoking guns: echoes and resonances}
%%%%%%%%%%%%%%%%%%%%%%%%%%%%%%%%%%%%%%%%%%%%%%%%%%%%
For binaries composed of ClePhOs, the GWs generated during inspiral and merger is expected to be very similar to
those by a corresponding BH binary with the same mass and spin. The arguments were detailed in Section~\ref{sec:stage}.
However, clear distinctive features appear due to the absence of a horizon. These features are:

\begin{itemize}
\item The appearance of late-time echoes in the waveforms~\cite{Cardoso:2016rao,Cardoso:2016oxy,Price:2017cjr,Nakano:2017fvh}. 
After the merger, ClePhOs give rise to two different ringdown signals: the first stage is dominated by the photosphere modes,
and is indistinguishable from a BH ringdown. However, after a time $\tau_{\rm echo}\sim M|\log\epsilon|$ (see Section \ref{sec:qnms}), the waves
trapped in the ``photosphere+ClePhO'' cavity start leaking out as echoes of the main burst. This is a smoking-gun for new physics, potentially reaching microscopic or even Planckian corrections at the horizon scale~\cite{Cardoso:2016rao,Cardoso:2016oxy}. 

Strategies to dig GW signals containing ``echoes'' out of noise are not fully under control, first efforts are underway~\cite{Echoes:datastrategies1,Echoes:datastrategies2}. 
The ability to detect such signals depends on how much energy is converted from the main burst into echoes (i.e., on the relative amplitude between the first echo and the prompt ringdown signal in Fig.~\ref{fig:ringdown}).
Define the ratio of energies to be $\gamma_{\rm echo}$. Then, the signal-to-noise ratio $\rho$ necessary for echoes to be detectable {\it separately} from the main burst, for a detection threshold of $\rho=8$, is
%
%\be
$\rho_{\rm prompt\, ringdown}\gtrsim \frac{80}{\sqrt{\gamma_{\rm echo}(\%)}}$.
%\ee
%
In such case the first echo is detectable with a simple ringdown template (an exponentially damped sinusoid~\cite{Abbott:2009km}).
Space-based detectors will see prompt ringdown events with very large $\rho$~\cite{Berti:2016lat}. For $\gamma_{\rm echo}=20\%$, we estimate that the planned space mission LISA~\cite{Audley:2017drz} will see at least one event per year, even for the most pessimistic population synthesis models used to estimate the rates~\cite{Berti:2016lat}. The proposed Einstein telescope~\cite{Punturo:2010zz} or Voyager-like~\cite{Voyager} third-generation Earth-based detectors will also be able to distinguish ClePhOs from BHs with such simple-minded searches.
The event rates for LIGO are smaller, and more sophisticated searches need to be implemented. Preliminary analysis of GW data using the entire echoing sequence was reported recently~\cite{Abedi:2016hgu,Ashton:2016xff,Abedi:2017isz}. 
More detailed characterization of the echo waveform (e.g., using Green's function techniques~\cite{Mark:2017dnq} and accounting for spin effects~\cite{Nakano:2017fvh}) is necessary to reduce the systematics in data analysis.
In several models of ClePhOs, $\gamma_{\rm echo}$ is large enough to produce effects that --~also owing to the logarithmic dependence of $\tau_{\rm echo}$ for all models known so far~-- are within reach of near-future GW detectors, even if the ClePhO corrections occurs at the Planck scale~\cite{Cardoso:2016rao,Cardoso:2016oxy}. This is a truly remarkably prospect. As the sensitivity of GW detectors increases, the absence of echoes might be used to {\rm rule out} many quantum-gravity models completely, and to push tests of gravity closer and closer to the horizon scale, as now routinely done for other cornerstones of GR, e.g. in tests of the equivalence principle~\cite{Will:2005va,Berti:2015itd}. 

\item Model-dependent fluid modes are also excited. Due to redshift effects, these will presumably play a subdominant role in the GW signal.

\item Certain models of ECOs arise naturally in effective theories with extra gravitational degrees of freedom~\cite{Mottola:2016mpl}. In such case, the detection of extra polarizations (as achievable in the future with a global network of interferometers) might provide evidence for new physics at the horizon scale.

\item Resonant mode excitation during inspiral. The echoes at late times are just vibrations of the ClePhO. These vibrations have relatively low frequency and can, in principle, also be excited during the inspiral process itself, leading to resonances in the motion as a further clear-cut signal of new physics~\cite{Pani:2010em,Macedo:2013qea,Macedo:2013jja}. 

\item Spin-mass distribution of compact objects skewed towards low spin, across the mass range. The development of the ergoregion instability depletes angular momentum from spinning ClePhOs, independently of their mass. Although the effectiveness of such process is not fully understood, it would lead to slowly-spinning objects as a final state.

\item  It is also possible that, at variance with the BH case, the merger of two ClePhOs can be followed by a burst of EM radiation associated with the presence of high-density matter during the collision. Although such emission might be strongly redshifted, searches for EM counterparts of candidate BH mergers can provide another distinctive signature of new physics at the horizon scale.
\end{itemize}

%%%%%%%%%%%%%%%%%%%%%%%%%%%%%%%%%%%%%%%%%%%%%%%%%%%%%%%%%%%%%%%%%%%%%%%%%%%%%%%%%%%%%%%%%%%%
\subsection{Precision physics: QNMs, tidal deformability, heating, and multipolar structure}
%%%%%%%%%%%%%%%%%%%%%%%%%%%%%%%%%%%%%%%%%%%%%%%%%%%%%%%%%%%%%%%%%%%%%%%%%%%%%%%%%%%%%%%%%%%%
The GW astronomy era will also gradually open the door to precision physics, for which smoking signs may not be necessary to test new physics.

\begin{itemize}
\item If the product of the merger is an ECO but not a ClePhO, it will simply vibrate differently from a BH.
Thus, precise measurements of the ringdown frequencies and damping times allow one to test whether or not the object is a BH~\cite{Berti:2005ys,Berti:2006qt}.
Such tests are in principle feasible for wide classes of objects, including boson stars~\cite{Berti:2006qt,Macedo:2013qea,Macedo:2013jja}, gravastars~\cite{Pani:2009ss,Mazur:2015kia,Chirenti:2016hzd}, wormholes~\cite{Konoplya:2016hmd,Nandi:2016uzg}, or other quantum-corrected objects~\cite{Barcelo:2017lnx,Brustein:2017koc}.

\item The merger phase can also provide information about the nature of the coalescing objects. ECOs which are not sufficiently compact will likely display a merger phase resembling that of a neutron-star merger rather than a BH merger~\cite{Cardoso:2016oxy,Bezares:2017mzk}. The situation for ClePhOs is unclear since no simulations of a full coalescence are available.

\item The absence of a horizon affects the way in which the inspiral stage proceeds. In particular, three different features may play a role, all of which can be used to test the BH-nature of the objects. 
\begin{itemize}

\item  \emph{No tidal heating for ECOs.} Horizons absorb incoming high frequency radiation, and serve as sinks or amplifiers for low-frequency radiation able to tunnel in (see Section~\ref{Sec:tides}).
UCOs and ClePhOs, on the other hand, are not expected to absorb any significant amount of GWs. Thus, a ``null-hypothesis'' test consists on using the phase of GWs to measure absorption or amplification at the surface of the objects~\cite{Maselli:2017cmm}. LISA-type GW detectors~\cite{Audley:2017drz} will place stringent tests on this property, potentially reaching Planck scales near the horizon and beyond~\cite{Maselli:2017cmm}.

\item  \emph{Nonzero tidal Love numbers for ECOs.} In a binary, the gravitational pull of one object deforms its companion, inducing a quadrupole moment proportional to the tidal field.
The tidal deformability is encoded in the Love numbers, and the consequent modification of the dynamics can be directly translated into GW phase evolution at higher-order in the post-Newtonian expansion~\cite{Flanagan:2007ix}. 
It turns out that the tidal Love numbers of a BH are zero~\cite{Binnington:2009bb,Damour:2009vw,Fang:2005qq,Gurlebeck:2015xpa,Poisson:2014gka,Pani:2015hfa}, allowing again for null tests. By devising these tests, existing and upcoming detectors can rule out or strongly constraint boson stars~\cite{Wade:2013hoa,Cardoso:2017cfl,Maselli:2017cmm,Sennett:2017etc} or even generic ClePhOs~\cite{Cardoso:2017cfl,Maselli:2017cmm}.

\item \emph{Different multipolar structure.} Spinning objects in a binary are also expected to possess multipole moments that differ from those of the Kerr geometry. This property is true at least for UCOs and for less compact objects. The impact of the multipolar structure on the GW phase will allow one to estimate and constrain possible deviations from the Kerr geometry~\cite{Krishnendu:2017shb}.

\end{itemize}

\end{itemize}

\newpage
%%%%%%%%%%%%%%%%%%%%%%%%%%%%%%%%%%%%%%%%%%%%%%%%%%%%%%%%%%%%%%%%%%%%%%%%%%%%%%
\section{Summary}
%%%%%%%%%%%%%%%%%%%%%%%%%%%%%%%%%%%%%%%%%%%%%%%%%%%%%%%%%%%%%%%%%%%%%%%%%%%%%%

% 
\hspace{1.8cm}
\parbox{0.8\textwidth}{{\small 
\noindent {\it ``The important thing is not to stop questioning. Curiosity has its own reason for existing.''}
% \vskip 2mm
\begin{flushright}
Albert Einstein, From the memoirs of William Miller, an editor, quoted in Life magazine, May 2, 1955; Expanded, p. 281 
\end{flushright}
}
}

\vskip 1cm

Thomson's atomic model was carefully constructed, and tested theoretically for inconsistencies\footnote{It is amusing that some alternative models -- which we now know to be closer to Rutherford's model -- were studied, shown to be unstable and therefore ruled out~\cite{1904Natur..69..437S}.}. Rutherford's incursion into scattering of $\alpha$ particles were not meant to disprove the model, they were aimed at testing its accuracy. According to Marsden, ``...it was one of those 'hunches' that perhaps some effect might be observed, and that in any case that neighbouring territory of this Tom Tiddler's ground might be explored by reconnaissance. Rutherford was ever ready to meet the unexpected and exploit it, where favourable, but he also knew when to stop on such excursions'' \cite{Birks}.

All the current observational evidence gathered around massive, compact and dark objects is compatible with the BH hypothesis.
There is no equally-well-motivated alternative, satisfying known laws of physics and composed of ordinary matter, which is compatible with those same observations.
Despite this, there are long-standing problems associated with horizons and singularities, which hint at some inconsistency between classical gravity and quantum mechanics at the scale of the horizon. Furthermore, since the majority of the matter in our Universe is unknown, we cannot exclude that dark compact objects made of exotic matter are lurking in the cosmos.
Even, and if, these issues are resolved
within classical physics, some outstanding questions remain. What evidence do we have that compact objects are BHs?
How deep into the potential can observations probe, and up to where are they still compatible with the current paradigm?
At one hand's reach, we have the possibility to dig deeper and deeper into the strong-field region of compact objects, {\it for free}. 
As the sensitivity of EM and GW detectors increases, so will our ability to probe 
regions of ever increasing redshift.
Perhaps the strong-field region of gravity holds the same surprises that the strong-field EM region did?

\vskip 2mm
\newpage
%%%%%%%%%%%%%%%%%%%%%%%%%%%%%%%%%%%%%%%%%%%%%%%%%%%%%%%%%%%%%%%%%%%%%%%%%%%%%%%
\noindent{\bf{\em Acknowledgments.}}
% %%%%%%%%%%%%%%%%%%%%%%%%%%%%%%%%%%%%%%%%%%%%%%%%%%%%%%%%%%%%%%%%%%%%%%%%%%%%%%
% \begin{acknowledgments}
%
We are indebted to Carlos Barcel\'o, Silke Britzen, Ram Brustein, Ra\'ul Carballo-Rubio, Bob Holdom, Luis Garay, Marios Karouzos, Gaurav Khanna, Joe Keir, Kostas Kokkotas, Claus Laemmerzahl, Jos\'e Lemos, Caio Macedo, Samir Mathur, Emil Mottola, Ken-ichi Nakao, Richard Price, Ana Sousa, Bert Vercnocke, Frederic Vincent, Sebastian Voelkel and Aaron Zimmerman for providing detailed feedback, useful references and for suggesting corrections to an earlier version of the manuscript.
We thank also the many other colleagues who provided constructive criticism and comments, including all the participants 
of the {\it Models of Gravity: Black Holes, Neutron Stars and the structure of space-time} meeting in Oldenburg,
of the {\it Foundations of the theory of Gravitational Waves} meeting in Stockholm, 
of the {\it Gravitational Waves and Cosmology} workshop at DESY,
of the {\it Workshop on Modern aspects of Gravity and Cosmology} in Orsay,
of the WE-Heraeus-Seminar on {\it Do Black Holes Exist? - The Physics and Philosophy of Black Holes} in Bad Honnef,
of the {\it GW161212: The Universe through gravitational waves} workshop at the Simons Center,
and of the {\it Quantum Vacuum and Gravitation: Testing General Relativity in Cosmology} workshop at the MITP.
V.C. acknowledges financial support provided under the European Union's H2020 ERC Consolidator Grant ``Matter and strong-field gravity: New frontiers in Einstein's theory'' grant agreement no. MaGRaTh--646597.
Research at Perimeter Institute is supported by the Government of Canada through Industry Canada and by the Province of Ontario through the Ministry of Economic Development $\&$
Innovation.
This article is based upon work from COST Action CA16104 ``GWverse'', and MP1304 ``NewCompstar'' supported by COST (European Cooperation in Science and Technology).
This work was partially supported by FCT-Portugal through the project IF/00293/2013, by the H2020-MSCA-RISE-2015 Grant No. StronGrHEP-690904.
% \end{acknowledgments}
%%%%%%%%%%%%%%%%%%%%%%%%%%%%%%%%%%%%%%%%%%%%%%%%%%%%%%%%%%%%%%%%%%%%%%%%%%%%%%

\bibliographystyle{utphys}

\bibliography{Ref}

\end{document}